\documentclass[a4paper,11pt]{article}
\usepackage[noblocks]{authblk}
\usepackage{times}
\usepackage{a4wide}

\title{Hierarchical Bond Graph Modelling of Biochemical Networks} 
\author{Peter J. Gawthrop\footnote{Corresponding author}}
 \affil{
   Systems Biology Laboratory,
   Melbourne School of Engineering,
   University of Melbourne,
   Victoria 3010, Australia.
\textbf{peter.gawthrop@unimelb.edu.au}}

\author{Joseph Cursons}
 \affil{
   Systems Biology Laboratory,
   Melbourne School of Engineering,
   University of Melbourne,
   Victoria 3010, Australia.
   \authorcr 
   ARC Centre of Excellence in Convergent Bio-Nano Science,
   Melbourne School of Engineering,
   University of Melbourne,
   Victoria 3010, Australia.
}

\author{Edmund J. Crampin}
\affil{
   Systems Biology Laboratory,
   Melbourne School of Engineering,
   University of Melbourne,
   Victoria 3010, Australia.
   \authorcr 
   School of Mathematics and Statistics,
   University of Melbourne,
   Victoria 3010, Australia.
   \authorcr 
   School of Medicine,
   University of Melbourne,
   Victoria 3010, Australia.
  \authorcr 
   ARC Centre of Excellence in Convergent Bio-Nano Science,
   Melbourne School of Engineering,
   University of Melbourne,
   Victoria 3010, Australia.
}

\usepackage{enumitem}
\setlist[itemize]{leftmargin=*}
\setlist[enumerate]{leftmargin=*}
\setlist[description]{leftmargin=*}

\usepackage{amsmath,amssymb,amsbsy,amscd,amstext,mathtools,amsxtra}
\newcommand{\lb}{\left (}
\newcommand{\rb}{\right )}
\newcommand{\diag}{\text{diag }}
\newcommand{\Ln}{\text{\bf Ln }}
\newcommand{\Exp}{\text{\bf Exp }}

\newcommand{\KK}{\boldsymbol{K}}
\newcommand{\kk}{\boldsymbol{k}}

\newcommand{\kkappa}{\boldsymbol{\kappa}}

\newcommand{\Nf}{{N^f}}
\newcommand{\Nr}{{N^r}}

\newcommand{\dX}{{\dot{X}}}
\newcommand{\dx}{{\dot{x}}}

\newcommand{\hp}{\circ}         

\newcommand{\vbond}{{\Huge $\rightharpoondown$}}
\newcommand{\RT}{RT}

\newcommand{\bK}{K^M}

\newcommand{\Vt}{{\tilde{V}}}

\newcommand{\reacul}[2]{
  {\; \xrightleftharpoons[#2]{#1} \;}
}

\newcommand{\reacu}[1]{
  \reacul{#1}{}
}

\newcommand{\reac}{
  \reacu{}
}

\newcommand{\BG}[1]{\text{\sffamily\textbf{#1}}}

\newcommand{\C}{\BG{C }}

\renewcommand{\SS}{\BG{SS }}
\newcommand{\one}{\BG{1 }}
\newcommand{\zero}{\BG{0 }}

\newcommand{\TF}{\BG{TF }}

\renewcommand{\Re}{\BG{Re }}

 \usepackage[numbers,compress]{natbib}
\usepackage{graphicx,subfigure,fancybox,color}

\newcommand{\Fig}[2]{
 \includegraphics[width=#2\linewidth]{#1.pdf}
}

\newcommand{\SubFig}[3]{
 \subfigure[#2]{
   \includegraphics[width=#3\linewidth]{#1.pdf}
   \label{subfig:#1}
 }
}

\newcommand{\SubModel}[2]{
\begin{figure}[htbp]
  \centering
  \Fig{#1_abg}{#2}
  \caption{Submodel: \Bg{#1}}
  \label{fig:#1}
\end{figure}
}

\newcommand{\ire}[2]{
\ar@^{>}@<0.3ex>[#1]^{#2}
}

\newcommand{\Bg}[1]{\text{\sffamily\textbf{#1}}}

\usepackage[pagebackref]{hyperref}
\hypersetup{colorlinks=true,citecolor=blue,linkcolor=blue,urlcolor=blue}
\usepackage{doi}
\usepackage{url}

\begin{document}
\maketitle
 \begin{abstract}
   The bond graph approach to modelling biochemical networks is
   extended to allow hierarchical construction of complex models from
   simpler components. This is made possible by representing the
   simpler components as thermodynamically open systems exchanging
   mass and energy via ports. A key feature of this approach is that
   the resultant models are \emph{robustly} thermodynamically
   compliant: the thermodynamic compliance is \emph{not} dependent on
   precise numerical values of parameters. Moreover, the models are
   \emph{reusable} due to the well-defined interface provided by the
   energy ports.

   To extract bond graph model parameters from
   parameters found in the literature, general and compact
   formulae are developed to relate free-energy constants and equilibrium
   constants. The existence and uniqueness of solutions is considered
   in terms of fundamental properties of stoichiometric matrices.

   The approach is illustrated by building a hierarchical bond graph
   model of glycogenolysis in skeletal muscle.
   






 \end{abstract}

\maketitle
\newpage
\tableofcontents
\newpage
\section{Introduction}
\label{sec:intro}

  A central aim of Systems Biology is to develop computational models
  for simulation and analysis of complex biological systems.  The
  scale of this challenge is well illustrated by two
  examples. Firstly, the recent whole-cell model of the pathogenic
  bacterium \emph{mycoplasma genitalium} \citep{KarSanMac12} which the
  authors claim accounts for the dynamics over the cell cycle of every
  gene product. Even for this relatively simple organism, the model
  was developed in the form of many submodels describing different cellular
  processes such as transcription, translation and metabolism; each
  submodel required a \emph{different} mathematical description and
  simulation approach, which were coupled together at simulation time.
  Secondly the Virtual Physiological Human project
  \citep{KohNob09,HunChaCov13} ``is a concerted effort to explain how
  each and every component in the body, from the scale of molecules up
  to organ systems and beyond, works as part of the integrated
  whole''\footnote{\url{http://physiomeproject.org/}}.

  In both examples, a key challenge is to devise modelling strategies and
  frameworks which not only handle complexity but also unite different
  aspects of cell biology, for example: cellular metabolism,
  signalling and electrophysiology. Such systems involve diverse
  physicochemical domains such as biochemical, chemoelectric and
  chemomechanical.
  There is therefore a need for a modelling approach to large
  multi-domain systems. As discussed below, the bond graph modelling
  formalism has long been used to model multi-domain engineering
  systems; and recently it has been applied to components of
  biochemical systems \citep{GawCra14}.  The purpose of this paper is
  extend the modelling of biochemical \emph{components} to the
  modelling of biochemical \emph{systems} by developing a hierarchical
  bond graph representation of biochemical systems.

As detailed in \citet{GawCra14}, thermodynamic aspects of
chemical reactions have a long history in the Physical Chemistry
literature. The roles of chemical potential and Gibb's
free energy in biochemical system analysis is described in
the textbooks of: \citet{Hil89}, \citet{BeaQia10} and
\citet{KeeSne09}.
  Indeed, if a computational model of a biochemical system is to be
  securely founded in the underlying chemistry, it must fully
  comply with the fundamental principles of physical chemistry as, for
  example laid out by \citet{AtkPau11}.

%
Thermodynamic compliance for a set of rate equations has been
considered in some detail by a number of authors
\cite{HenBroHat07,EdeGil07,LieUhlKli10} allowing
this theory to be applied over preexisting models and providing 
rules to examine new models and ensure that they are constructed 
to be compliant. However, when rules such as the Haldane constraint
are used to constrain a set of numerical parameter values, 
numerical errors may still destroy thermodynamical compliance.
%
The bond graph approach to modelling dynamical systems is well
established - a comprehensive account is given in the textbooks of
\citet{GawSmi96}, \citet{Bor11} and \citet{KarMarRos12} and a tutorial
introduction for control engineers is given by \citet{GawBev07}. It
has been applied to chemical reaction networks by \citet{OstPerKat71,OstPerKat73}
and to chemical reactions by \citet{Cel91}, \citet{GreCel12} and
\citet{GawCra14}.  \citet{ThoAtl85} discuss ``osmosis as chemical
reaction through a membrane'', \citet{LefLefCou99} model cardiac
muscle using the bond graph approach  and \citet{DiaPic11} discuss
the application of bond graphs to biological system in general and
muscle models in particular.  As illustrated in this paper, the bond
graph method provides an alternative approach to ensuring
thermodynamic compliance: a biochemical reaction network built out of
bond graph components \emph{automatically} ensures thermodynamic
compliance even if numerical values are not correct. In this sense,
the bond graph approach is \emph{robustly} thermodynamically compliant
(RTC).
  Conversely, the discipline imposed by bond graph modelling can alert
  the modeller to errors and inconsistencies that may otherwise have
  escaped attention. This is illustrated in this paper in the context
  of revising a well-established model of glycogenolysis
  \citep{LamKus02}.

Thermodynamics distinguishes between: \emph{open systems} where
energy and matter can cross the system boundary; \emph{closed systems}
where energy, but not matter, can cross the system boundary; and
\emph{isolated systems} where neither energy nor matter can cross the system
boundary \citep{AtkPau11}. Living organisms and their subsystems are
open as both matter and energy cross the boundaries, thus,
mathematical models of living organisms and their subsystems should
have explicit formulation as an open system. In the bond graph approach
used for this paper, open subsystem models explicitly contain \emph{ports}
through which matter and energy can flow.

The need for a systems level approach has been termed ``understanding
the parts in terms of the whole'' by \citet{CorCarLet04} and the need
for mathematical methods to integrate the range of cellular processes
has been emphasised by \citet{GonBucRyl13}.
Building such system-level models from components is simplified if
preexisting validated models can be \emph{reused} and combined in new
models.
  A number of standards have been established to distribute re-usable
  and consistent models of biochemical networks \citep{HucFin05},
  including the mark-up languages SBML \citep{HucFinSau03} and CellML
  \citep{CooHunCra08}, and reporting guidelines such as the Minimum
  information requested in the annotation of biochemical models
  (MIRIAM) \citep{NovFinHuc05}. A number of models described with SBML
  and CellML have been published, and many of these are stored within
  the BioModels Database \citep{LiDonRod10} and the Physiome
  Repository \citep{HunBor03}.  More recently developed, SED-ML
  \citep{WalAdaBer11} now allows ``simulated experiments'' with SBML
  and CellML models to be described.
  However, despite the increasing availability of models written using
  standardised formats, it is still a difficult and complex task to
  combine such models within a larger system
  \citep{TerNieCra08,NeaCooSmi14}.

In the context of a thermodynamically compliant reaction
network, model reuse requires that interconnecting models with preexisting
thermodynamical compliance produces a model where this property is
inherited. As discussed in this paper, bond graph models for open
systems contain energy ports which provide the required
interconnections thus allowing the models to be combined to produce a
higher-level model which is robustly thermodynamically compliant. 
%
This form of hierarchical modelling has been used in
the bond graph context by \citet{Cel92} and makes use of bond graph 
energy ports \citep{Cel91,KarMarRos12}.
As discussed in more detail by \citet{GawBev07} and later within this 
paper, the \emph{source sensor} (\SS) bond graph component, introduced 
by \citet{GawSmi92a}, provides a convenient notation for such energy
ports.
%
Because the system components are connected by energy ports,
interactions between the components are two-way, thus the distinction
between input and output is meaningless and a component's dynamic
properties depend on those components which are connected to it. Thus
the analogy with \emph{active} electronic components (\textit{e.g.} diodes and 
transistors) designed to avoid such
two-way interactions is not helpful here. Rather, the analogy is with
\emph{passive} electronic components (\textit{e.g.} resistors and capacitors).

\citet{GawCra14} also discuss how bond graph models are parameterised
such that parameters relating to thermodynamic quantities are clearly 
demarcated from reaction kinetics. However, this alternative
parameterisation leads to an important issue: the parameters required
by the bond graph model are not the same as the parameters normally
used to describe a biochemical reaction network. Thus parameters must
be converted from one form to another. This conversion is not trivial
for two reasons: the given parameters must be checked for
thermodynamic compliance for a solution to exist, and this solution
is not necessarily unique. The problem is solved in this paper and illustrated
using a specific example from the literature.

Complex systems may be difficult to comprehend theoretically, and the
practical derivation of properties is often tedious. These issues can be 
overcome to some extent by providing general representations of, and
general formulae relating to arbitrarily complex systems. The
stoichiometric matrix approach \citep{Pal06,Pal11} is one such
methodology; another approach is provided by \citet{SchRaoJay13} who
investigate the mathematical structure of balanced chemical reaction
networks governed by mass action kinetics and provide a powerful
mathematical notation. This paper provides a general representation of
an open biochemical reaction network described using bond graphs as well
as general formulae based on the stoichiometric matrix
\citep{Pal06,Pal11} approach and on the mathematical structure and
notation of \citet{SchRaoJay13}.
  As discussed by \citet[\S~3]{GawCra14}, the use of the bond graph
  concept of \emph{causality} visually and intuitively ties the
  mathematical concepts of the stoichiometric matrix and its subspaces
  (representing pathways and conserved moieties)  \citep{Pal06,Pal11}
  to the actual chemistry underlying the system as represented by the
  biochemical system bond graph.

One purpose of this paper is to introduce Systems Biologists to the
bond graph modelling approach. To this end, a metabolic pathway model
from the literature is re-implemented and reexamined from the bond
graph point of view.  In particular, \citet{LamKus02} consider ``a
computational model for glycogenolysis in skeletal muscle'' and
provide a detailed thermodynamically consistent model. This model has
been further embellished by 
\citet{VinRusPal10} who also provide a detailed comparison with
experimental data. \citet[Chapter 7]{Bea12} uses this model as an
example for the simulation of complex biochemical cellular systems,
and \citet{MosAlfMaj12} uses a similar model to examine the
interaction of the AKT pathway on metabolism. This paper uses the
\citet{LamKus02} model as an exemplar of how the bond graph approach
can be applied to build a hierarchical model which is RTC.

The outline of the paper is as follows.
\S~\ref{sec:closed-open-systems} discusses the modelling of open
thermodynamical systems using bond graphs and the bond graph notion of
an \emph{energy port};
\S~\ref{sec:simple-exampl-extern}
introduces the topic using a simple example,
\S~\ref{sec:general-case} gives the
general case and \S~\ref{sec:energy-flow}
focuses on energy flows.
Conversion of kinetic data into the form
required for bond graphs is considered in
\S~\ref{sec:conv-kinet-data}; in particular
\ref{sec:equil-const} examines the
relation between equilibrium constants and free-energy constants and
gives a general formula. Existence and uniqueness issues are discussed
and an alternative derivation of the Wegscheider conditions is given and
\S~\ref{sec:enzyme-catal-react}
discusses the reformulation of enzyme-catalysed reactions.
\S~\ref{sec:hier-modell} develops a hierarchical bond graph model
based on the well-established model of glycogenolysis by
\citep{LamKus02} to illustrate the concepts developed in this paper,
and \S~\ref{sec:conclusion} concludes the paper with suggested
directions for further work.

A virtual reference environment \citep{HurBudCra14} is available for
this paper at \url{https://sourceforge.net/projects/hbgm/}.

\paragraph{Notation}
Following \citet{SchRaoJay13}, we use the convenient notation $\Exp X$
to denote the vector whose $i$th element is the exponential of the
$i$th element of $X$ and $\Ln X$ to denote the vector whose $i$th
element is the natural logarithm of the $i$th element of
$X$. \citet{SchRaoJay13} also use the element-wise multiplication
operator ``$.$''; here, for clarity, we use the equivalent Hadamard or
Schur product with symbol ``$\hp$''.
The Hadamard or Schur operator $\hp$ \citep[Sec. 7.3]{Ber05}
corresponds to element-wise multiplication of two matrices which have the same
dimensions. This is the same as the $.*$ notation within matrix
manipulation software. Note that $K \hp X = X \hp K = (\diag K) X =
(\diag X) K$, where the notation ``$\diag X$'' is used for the 
matrix with elements of $X$ along the diagonal.

\section{Closed and Open systems}
\label{sec:closed-open-systems}
\begin{figure}[htbp]
  \centering 
  \SubFig{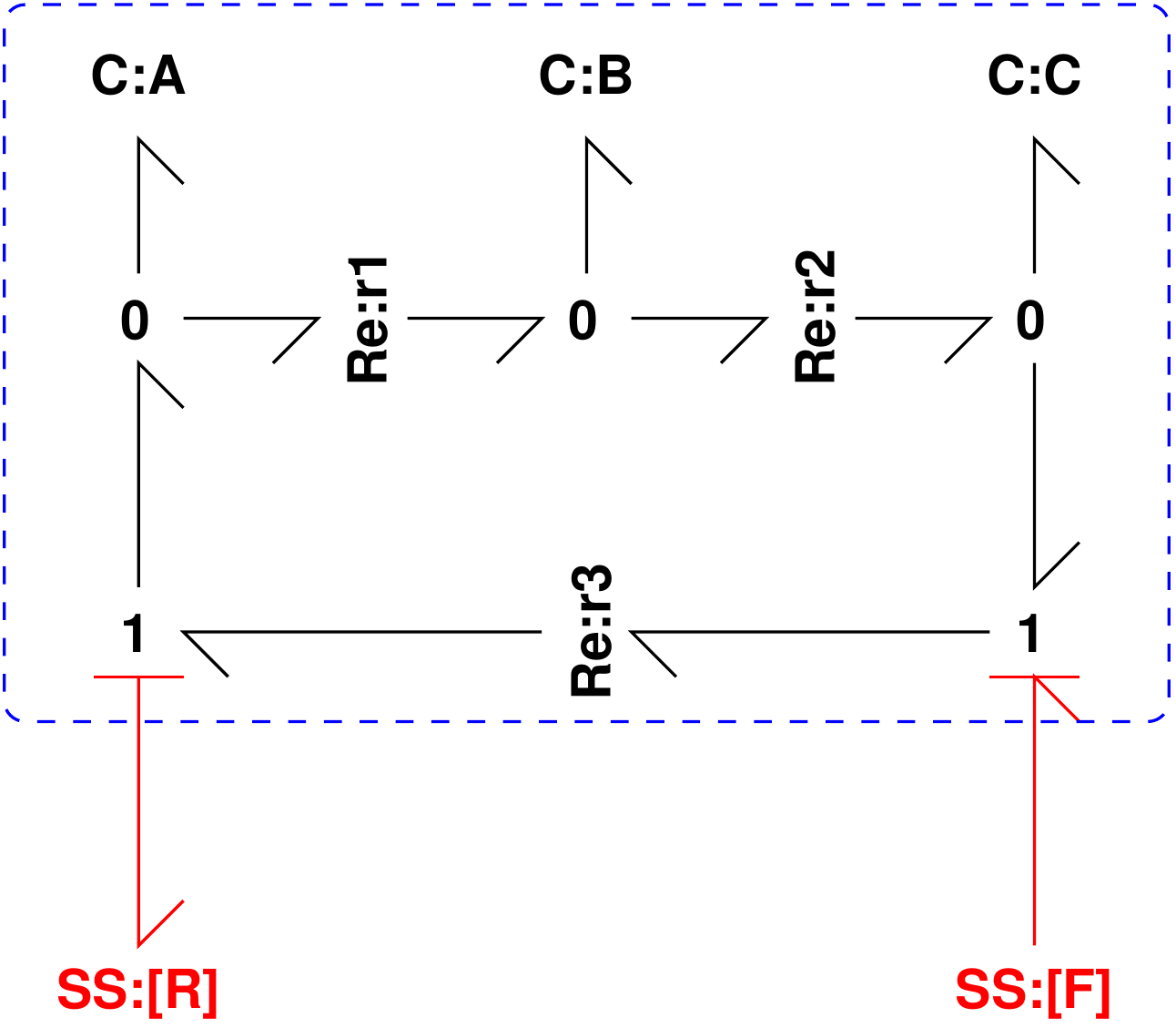}{Example: external efforts  $\mu_F$ and $\mu_R$.}{0.45}
  \SubFig{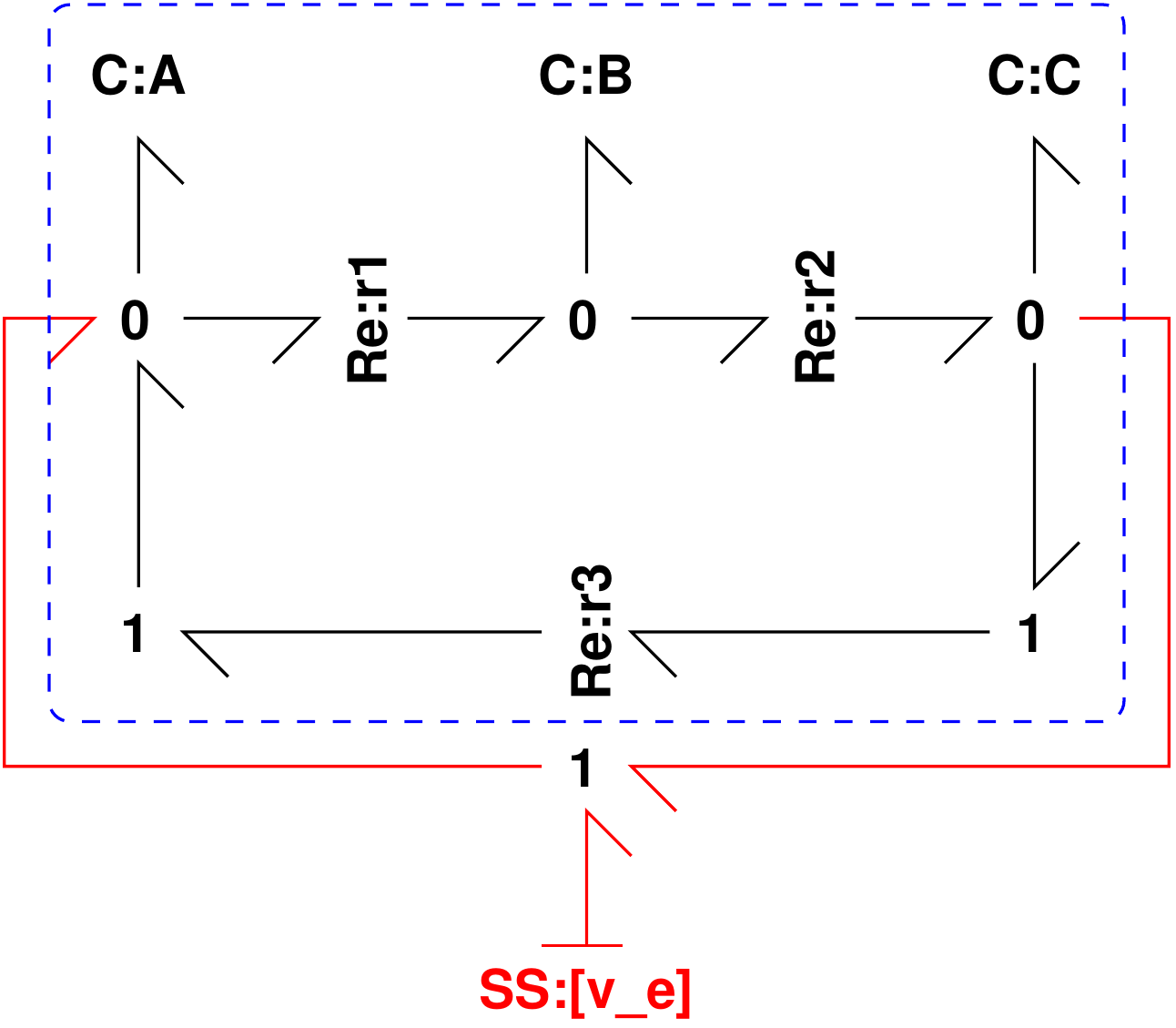}{Example: external flow $v_e$}{0.45}
  \SubFig{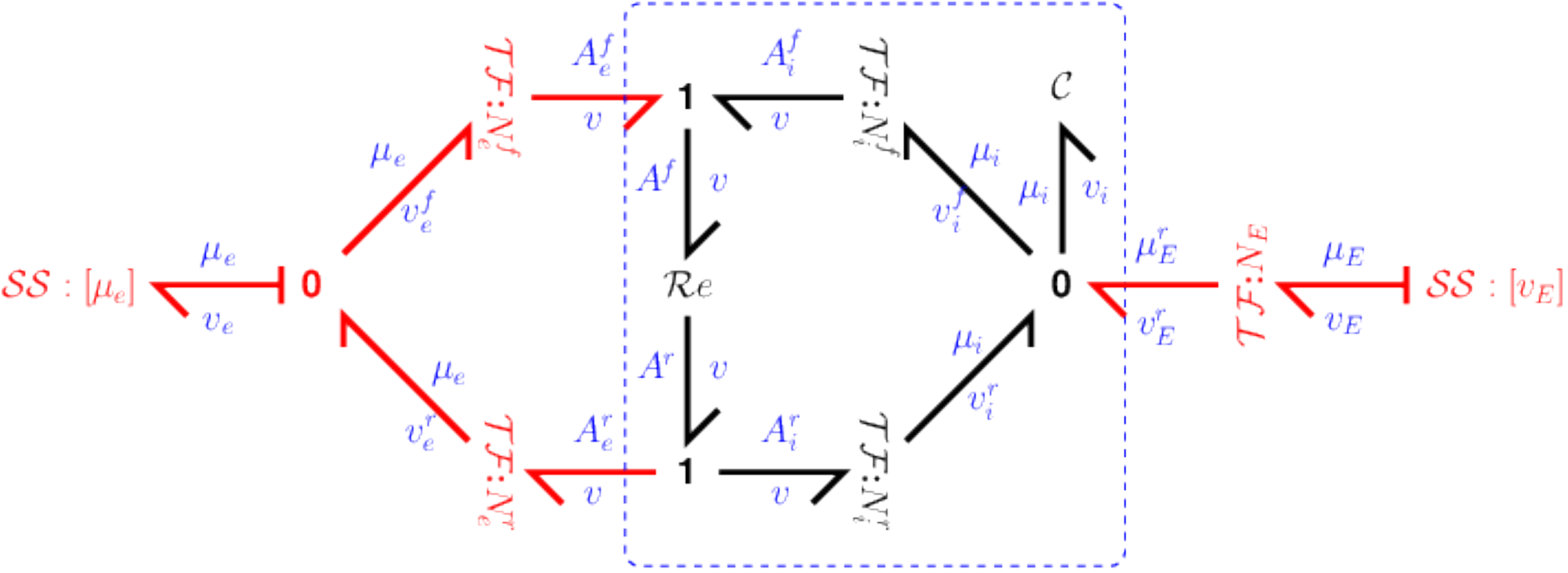}{General}{0.9}
  \caption{Closed \& open systems. (a) \& (b) Simple examples corresponding
    to reaction schemes \eqref{eq:ABCA_reac} and \eqref{eq:ABCAo_reac} with
    imposed concentrations and flows, respectively.
    The \Bg{C:A}, \Bg{C:B}, and \Bg{C:C} components accumulate the
    species $A$, $B$ and $C$; and the \Bg{Re:r1}, \Bg{Re:r2} and
    \Bg{Re:r3} represent reactions 1, 2 and 3. The \Bg{SS:[F]},
    \Bg{SS:[R]} and \Bg{SS:[v\_e]}
    components are the energy ports converting a closed system to an open one.
    The system inside the dashed box is a closed system; when
    \Bg{SS} components (representing energy ports) are added the
    system becomes open. (c) General case. The
    bond symbols
    $\rightharpoondown$ correspond to \emph{vectors} of bonds; 
    $\mathcal{C}$, $\mathcal{R}e$, $\mathcal{SS}$, \zero and \one
    symbols correspond to arrays of the associated components; and 
    $\mathcal{TF}$ components represent the intervening
    junction structure which transmits energy, comprising bonds, junctions
    and \TF components. $\mathcal{SS}:[\mu_e]$ represents a vector of
    energy ports imposing external chemical potentials;
    $\mathcal{SS}:[v_E]$ represents a vector of energy ports imposing
    external flows. In particular, with respect to (a) \& (b),
    $\mathcal{C}$ corresponds to the species represented by \Bg{C:A},
    \Bg{C:B}, and \Bg{C:C},
    $\mathcal{R}e$ corresponds to the reactions represented by \Bg{Re:r1}, \Bg{Re:r2} and
    \Bg{Re:r3} and
    $\mathcal{TF}:N^f_i$ and $\mathcal{TF}:N^r_i$ summarise the
    connections between reactions and species. 
    With respect to (a), $\mathcal{SS}:[\mu_e]$ corresponds to \Bg{SS:[F]},
    \Bg{SS:[R]}; with respect to (b), $\mathcal{SS}:[v_E]$ corresponds
    to \Bg{SS:[v\_e]}. 
    $\mathcal{TF}:N^f_e$ and $\mathcal{TF}:N^r_e$ summarise the
    connections between reactions and ports, and $\mathcal{TF}:N_E$
    between species and ports, in (a) \& (b).  }
 \label{fig:closed-open-systems}
\end{figure}
This section discusses the modelling of open thermodynamic systems
using bond graphs. In particular, it is shown how an otherwise closed
system becomes an open system by the addition of energy ports and the
standard bond graph components that represent such ports are introduced.
A simple motivational example is discussed in
\S~\ref{sec:simple-exampl-extern} and the
general case is discussed in
\S~\ref{sec:general-case}.

\subsection{Simple Example: open and closed systems}
\label{sec:simple-exampl-extern}
As described in \citet{GawCra14}, the reaction scheme:
\begin{xalignat}{3}
  A &\reacu{1} B &
  B &\reacu{2} C &
  C  &\reacu{3} A \label{eq:ABCA_reac}
\end{xalignat}
has the bond graph representation given within the dashed box of
Figures \ref{subfig:ABCAo_abg} and \ref{subfig:ABCAv_abg}. This is a
\emph{closed} system with respect to the three species $A$, $B$ and
$C$, thus the system equilibrium (where the quantities of $A$, $B$ and
$C$ are constant) corresponds to zero reaction flows.

In contrast, consider the reaction scheme:
\begin{xalignat}{3}
  A &\reacu{1} B &
  B &\reacu{2} C &
  C + F &\reacu{3} A + R \label{eq:ABCAo_reac}
\end{xalignat}
where the reactants $F$ and $R$ correspond to large external pools
with effectively fixed concentrations $X_f$ and $X_r$. This is
represented in Figure \ref{subfig:ABCAo_abg} where two instances of
the \emph{effort source} and \emph{flow sensor} components, \Bg{SS:[F]} and
\Bg{SS:[R]}, have been appended to the bond graph. Each \Bg{SS} imposes a
chemical potential (effort in bond graph parlance) upon the system and
inherits the flow on the corresponding \Bg{1} junction. With the sign 
convention shown, energy flows into the system through \Bg{SS:[F]} and
out of the system through \Bg{SS:[R]}. 

Although there is a net flow of energy into the system, in this
particular case there is no net flow of material. The two \Bg{1}
junctions are connected by an \Bg{Re} component thus the flows are
equal.

Further details are given within \citet{GawCra14} as to how the mass-action
kinetics of this closed system can be directly derived from the bond graph:
\begin{xalignat}{3}
\label{eq:intro:closed_x}
  \dx_a &=  v_3 - v_1&
  \dx_b &=  v_1 - v_2&
  \dx_c &=  v_2 - v_3
\end{xalignat}
where: $x_a$, $x_b$ and $x_c$ are the amounts of species $A$, $B$ and
$C$ in moles; and $v_1$, $v_2$ and $v_3$ are the molar flows associated
with reactions 1--3, given by:
\begin{xalignat}{3}
\label{eq:intro:closed_v}
  v_1 &= \kappa_1 \left( K_a x_a - K_b x_b \right)&
  v_2 &= \kappa_2 \left( K_b x_b - K_c x_c \right)&
  v_3 &= \kappa_3 \left( K_c x_c - K_a x_a \right)
\end{xalignat}

These linear equations can be written in the notation of \S~\ref{sec:intro}
using the stoichiometric matrix $N$:
\begin{xalignat}{3}
 \label{eq:simple_open_V}
  \dX &= N V&
  V &= -\kappa \hp \lb N^T K \hp X \rb&
  \text{where }N &=
  \begin{pmatrix}
    -1 & 0 & 1\\
    1 & -1 & 0\\
    0 & 1 & -1
  \end{pmatrix}
\end{xalignat}
with associated mass and flow vectors, and coefficient matrices:
\begin{xalignat}{4}
 \label{eq:simple_open_V_cont}
  X &=
  \begin{pmatrix}
    x_a\\x_b\\x_c
  \end{pmatrix}&
  V &=\begin{pmatrix}
    v_1\\v_2\\v_3
  \end{pmatrix}&
  \kappa &=
  \begin{pmatrix}
    \kappa_1\\
    \kappa_2\\
    \kappa_3
  \end{pmatrix}&
  K &= 
  \begin{pmatrix}
    K_a\\
    K_b\\
    K_c
  \end{pmatrix}
\end{xalignat}
%
The open system given by the reaction scheme (\ref{eq:ABCAo_reac}) and the bond graph of Figure
\ref{subfig:ABCAo_abg} is given by Equations (\ref{eq:intro:closed_x})
but the equation for $v_3$ of Equations (\ref{eq:intro:closed_v}) is replaced by:
\begin{equation}
\label{eq:simple_closed_3}
   v_3 = \kappa_1 \left( K_f x_f K_c x_c - K_r x_r K_a x_a \right)
\end{equation}
The simple linear expression \eqref{eq:simple_open_V} is also no longer
applicable, but as discussed in \S~\ref{sec:general-case}
it can be replaced in general terms by:
\begin{xalignat}{3}
 \label{eq:simple_open_V_1}
  \dX_i &= N_i V& 
  V &= \kappa \hp \lb  \Exp \Nf^T \Ln K \hp X - \Exp \Nr^T \Ln K \hp X  \rb&
  \text{where } N_i &= N^r_i - N^f_i 
\end{xalignat}
and where
\begin{xalignat}{4}
\label{eq:N_simple}
  X &=
  \begin{pmatrix}
    x_a\\x_b\\x_c\\x_f\\x_r
  \end{pmatrix}&
  K &= 
  \begin{pmatrix}
    K_a\\K_b\\K_c\\K_f\\K_r
  \end{pmatrix}&
   N^f &=
   \begin{pmatrix}
     N^f_i \\ N^f_e
   \end{pmatrix}& 
   N^r &=
   \begin{pmatrix}
     N^r_i \\ N^r_e
   \end{pmatrix}
\end{xalignat}
\begin{xalignat}{4}\label{eq:simple_N}
N^f_i &=
\begin{pmatrix}
  1 & 0 & 0\\
  0 & 1 & 0\\
  0 & 0 & 1
\end{pmatrix}
& N^r_i &= 
\begin{pmatrix}
  0 & 0 & 1\\
  1 & 0 & 0\\
  0 & 1 & 0
\end{pmatrix}&
N^f_e &=
\begin{pmatrix}
  0 & 0 & 1\\
  0 & 0 & 0
\end{pmatrix}
& N^r_e &= 
\begin{pmatrix}
  0 & 0 & 0\\
  0 & 0 & 1
\end{pmatrix}
\end{xalignat}
\begin{figure}[htbp]
  \centering
  \SubFig{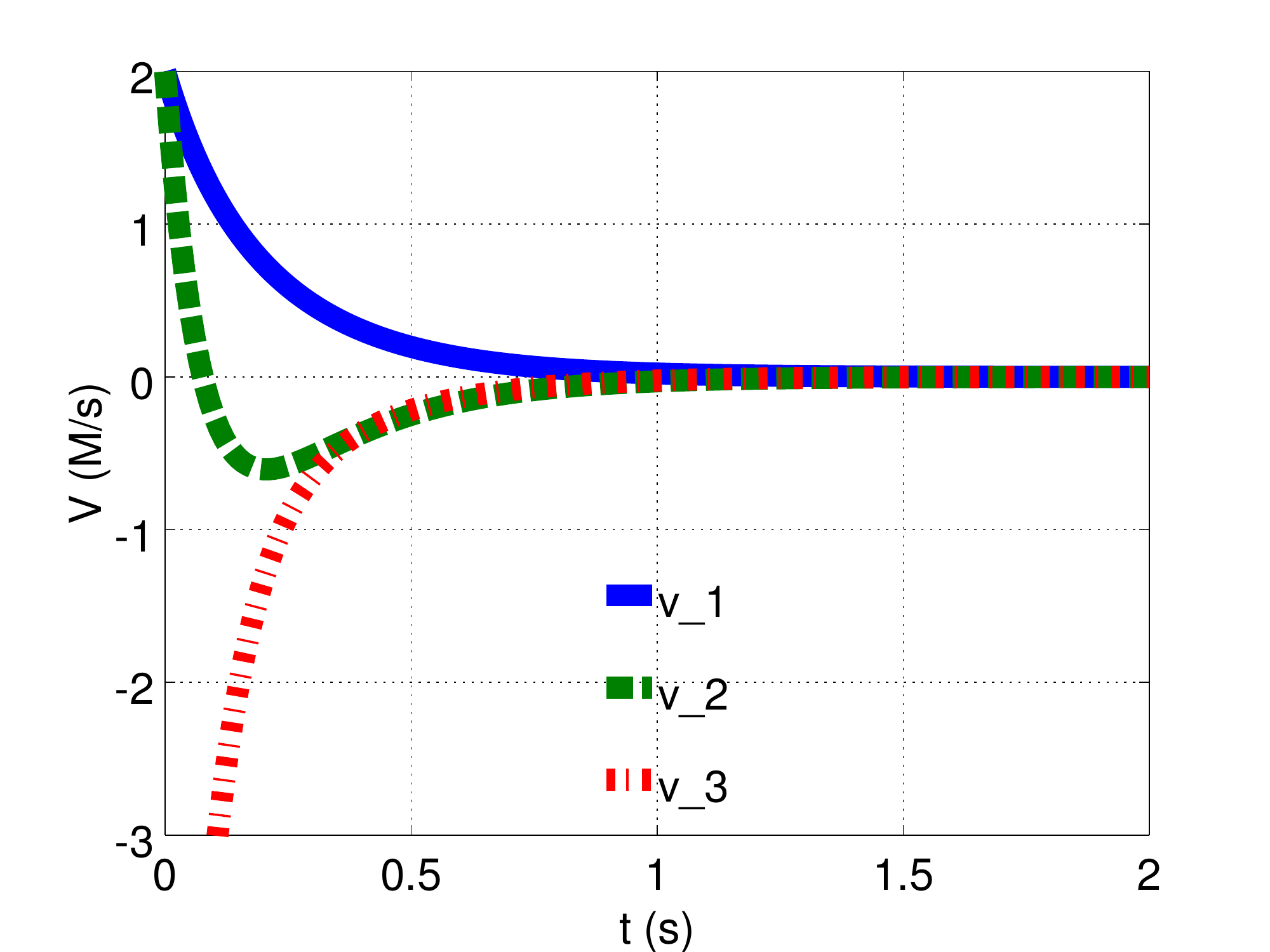}{Closed system: $V$}{0.45}
  \SubFig{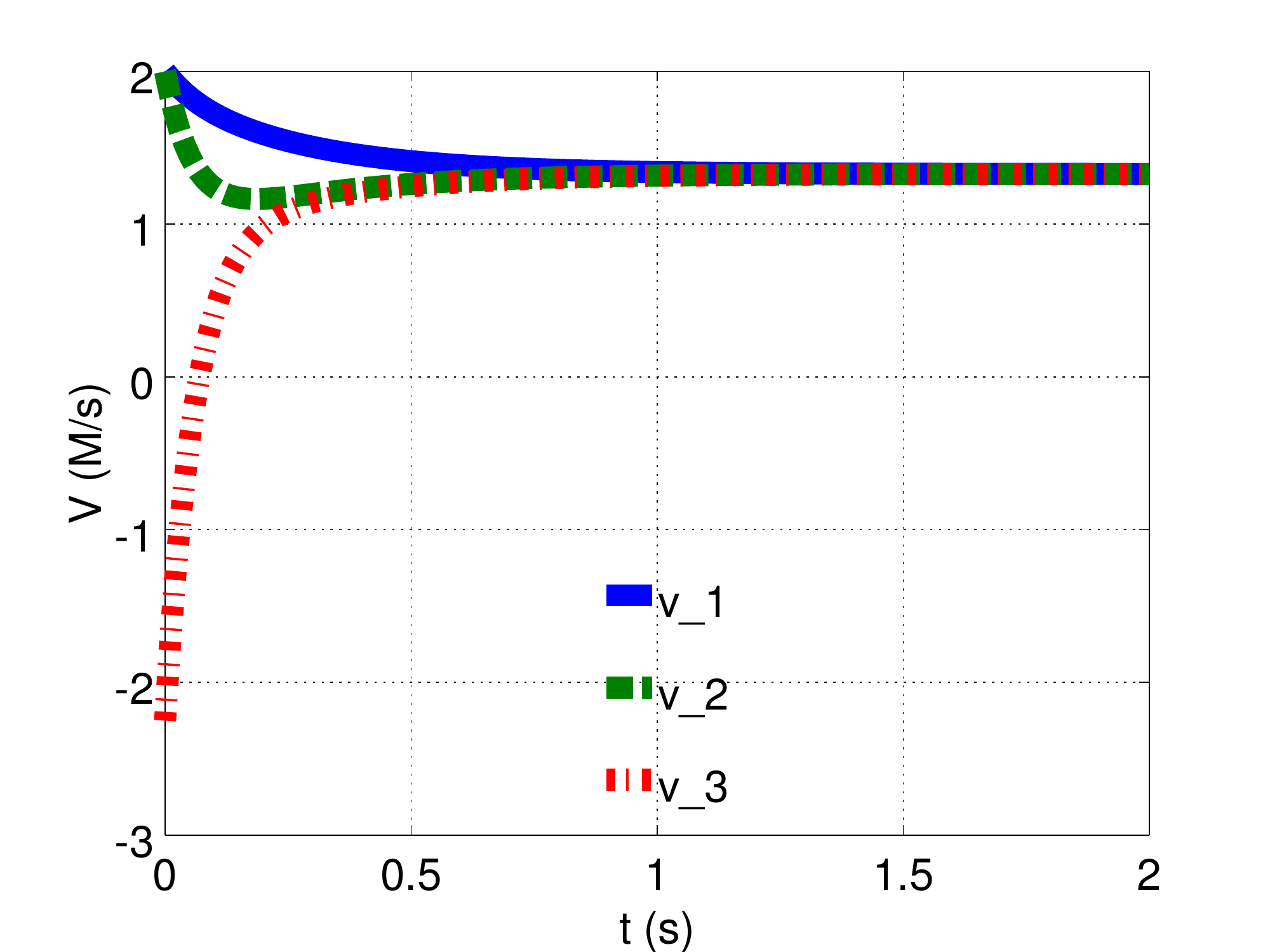}{Open system: $V$}{0.45}
  \SubFig{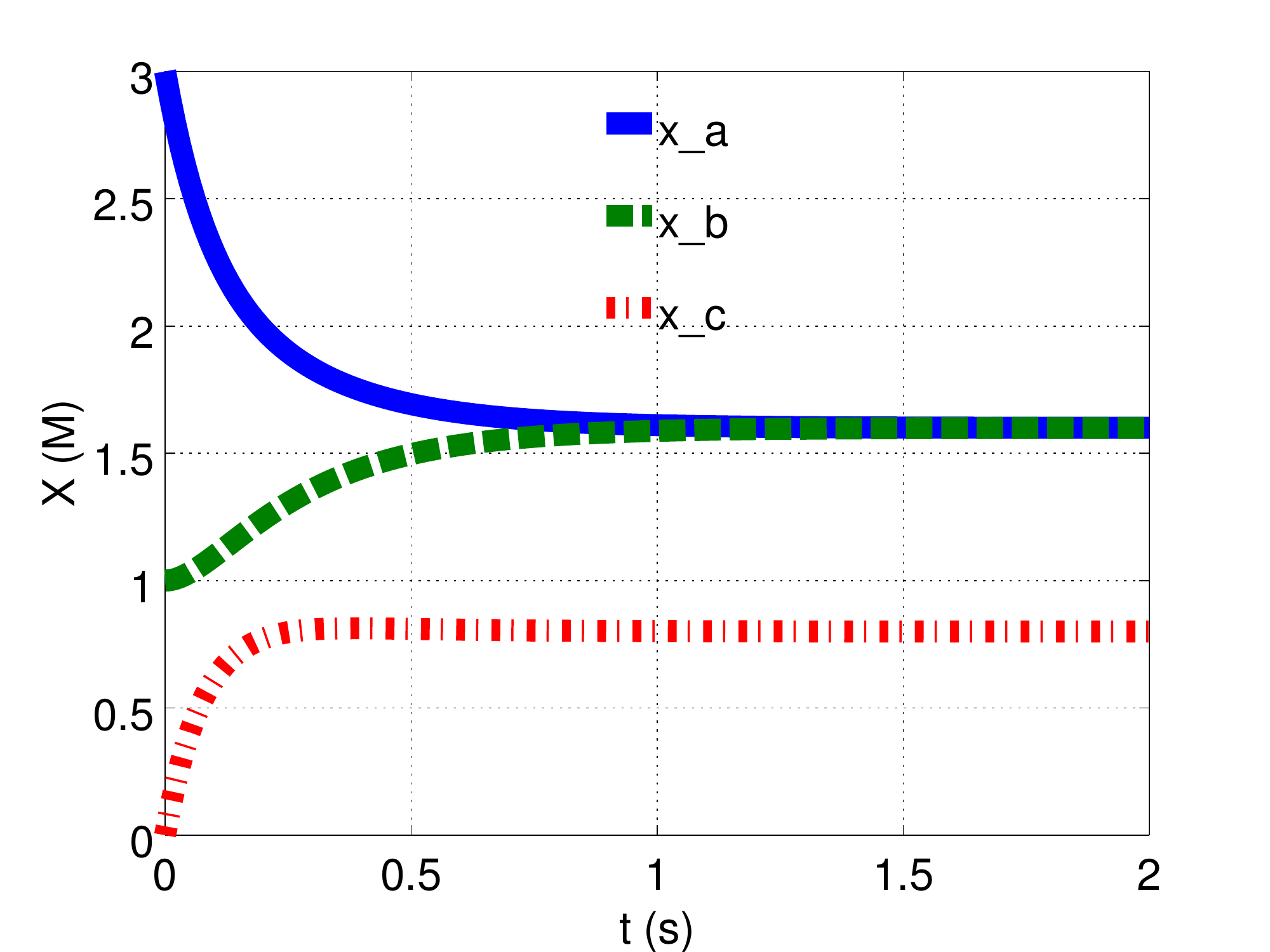}{Closed system: $X$}{0.45}
  \SubFig{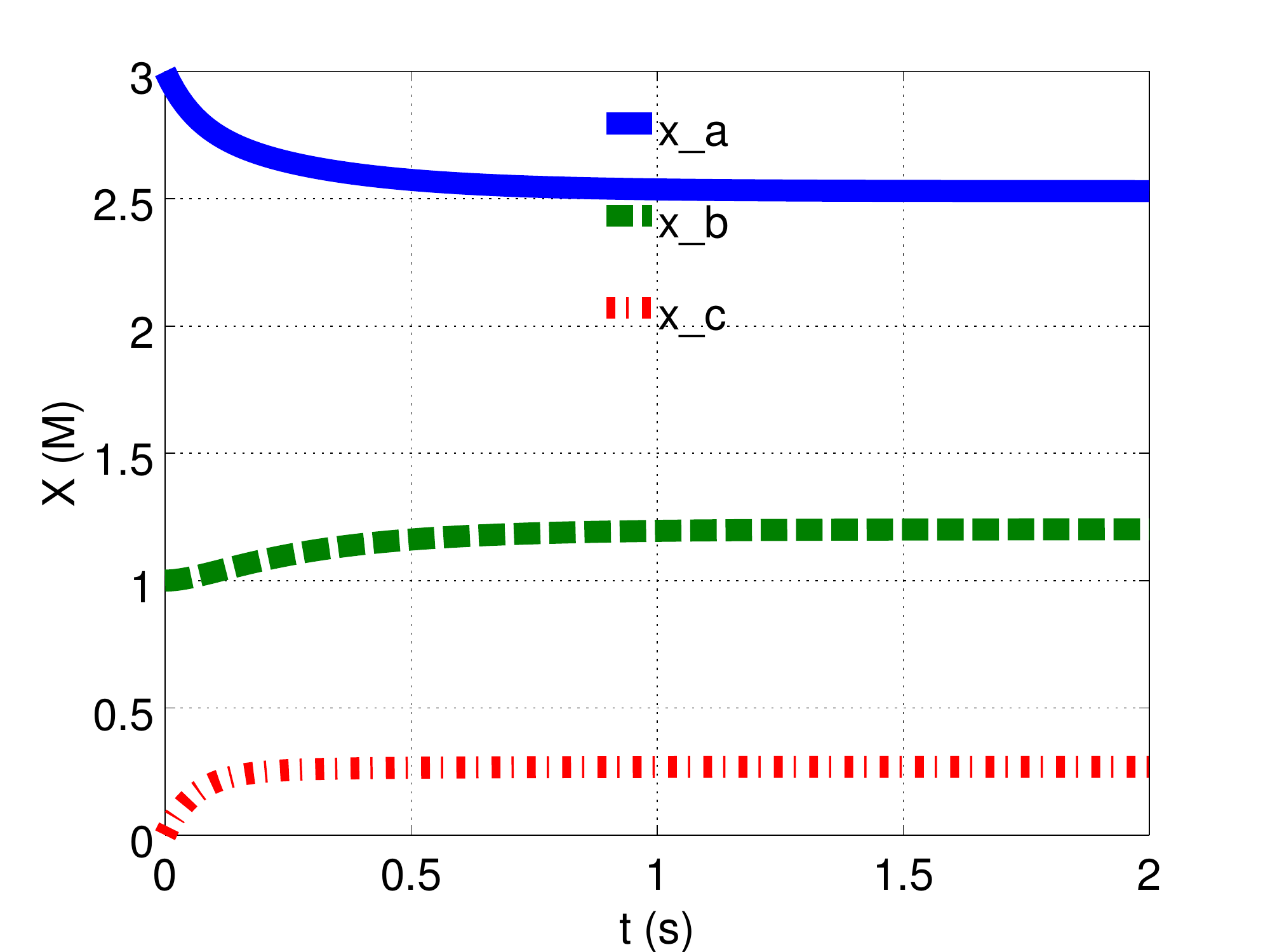}{Open system: $X$}{0.45}
  \caption{Simulation of closed \& open systems. The parameters are
    arbitrarily chosen as:
    $K_a=K_b=1$, $K_c=K_f =2$, $K_r=0.25$, $X_f=X_r=1$, $\kappa_1=1$,
    $\kappa_2=2$ \& $\kappa_3=3$.
    (a) As this is a
    closed system, the three reaction flows $v_1 \dots v_3$ become zero
    as time increases.
    (b) As this is an open system, the three reaction flows
    $v_1 \dots v_3$ do not become zero as the system has an external
    driver. However, as the states become constant, the three flows
    become equal.  
    (c) Although the closed system flows become zero the states do not, 
    as the system has a conserved moiety: the three states have a 
    constant sum $x_a+x_b+x_c=4$.
    (d) The three states tend to different values to those in (c); but
    the moiety is still conserved.  }
  \label{fig:sim_ABCAo}
\end{figure}
A simulation of this simple example is shown in Figure
\ref{fig:sim_ABCAo} in order to emphasise the main distinctions
between open and closed systems.




\subsection{General case}
\label{sec:general-case}
Figure \ref{subfig:Open_abg} is a generalisation of Figures
\ref{subfig:ABCAo_abg} and \ref{subfig:ABCAv_abg}, giving a
schematic representation of the structure for an open system.
With reference to Figure \ref{subfig:Open_abg}, the five
$\mathcal{TF}$ components represent the five flow transformations:
\begin{xalignat}{5}
  \label{eq:genTF}
  v^f_i &= N^f_i v &
  v^r_i &= N^r_i v &
  v^f_e &= N^f_e v &
  v^r_e &= N^r_e v &
  v^r_E &= N_E v_E 
\end{xalignat}
and thus the five matrices $N^f_i$, $N^r_i$, $N^f_e$, $N^r_e$ \& $N_E$
define the stoichiometry of the chemical network.

 

As emphasised by \citet{GawCra14} - bonds, \Bg{TF} components and
junctions all transport, but do not create or destroy, chemical energy. It follows
from this that:
\begin{xalignat}{4}
  \label{eq:conserve}
  {v}^T A^f_i &=  {v^f_i}^T \mu_i&
  {v}^T A^r_i &=  {v^r_i}^T \mu_i&
  {v}^T A^f_e &=  {v^f_e}^T \mu_e&
  {v}^T A^r_e &=  {v^r_e}^T \mu_e
\end{xalignat}
Using equations \eqref{eq:genTF}, equations \eqref{eq:conserve} become:
\begin{xalignat}{4}
  \label{eq:conserve_2}
  {v}^T A^f_i &= {v}^T{N^f_i}^T  \mu_i&
  {v}^T A^r_i &= {v}^T{N^r_i}^T \mu_i&
  {v}^T A^f_e &= {v}^T{N^f_e}^T \mu_e&
  {v}^T A^r_e &= {v}^T{N^r_e}^T \mu_e
\end{xalignat}
As the equations \eqref{eq:conserve_2} must be true for all $v$, it
follows that the affinities $A$ can be expressed in terms of chemical
potential $\mu$ as:
\begin{xalignat}{5}
  \label{eq:affinities}
   A^f_i &= {N^f_i}^T  \mu_i&
   A^r_i &= {N^r_i}^T \mu_i&
   A^f_e &= {N^f_e}^T \mu_e&
   A^r_e &= {N^r_e}^T \mu_e&
   \mu_E &= {N_E}^T \mu_i
\end{xalignat}
The fact that energy conservation implies Equation
\eqref{eq:affinities} is discussed by \citet{Cel91}.

With the notation given in \S~\ref{sec:intro}, the \emph{Marcelin} formula for reaction flows
(used by \citet[Eqn. (2.6)]{GawCra14} for the scalar case) may be
written
%
 as:
\begin{xalignat}{3}
 \label{eq:Marcelin}
 V &= \kappa \hp
 \lb V_0^+ - V_0^- \rb  + N_E v_E&
\text{where } V_0^+ &= \Exp{\frac{A^f}{RT}}&
\text{and } V_0^- &= \Exp{\frac{A^r}{RT}}
\end{xalignat}
Similarly, the formula for chemical potential $\mu$ of
\citet[Eqn. (2.3)]{GawCra14} in terms of the vector $K_i$ of
\emph{free-energy constants} is given by:
\begin{xalignat}{2}\label{eq:mu}
  \mu_i &=  RT \Ln  K_i \hp X_i & 
\text{where } K_i &= \Exp \frac{\mu_i^0}{RT}
\end{xalignat}
where $\mu_i^0$ is the vector of  standard chemical potentials for the
internal species.
It is also convenient to re-express the external chemical potentials $\mu_e$
as $\mu_e =  RT \Ln K_e \hp X_e$.
Combining Equations \eqref{eq:Marcelin}--~\eqref{eq:mu}, the
reaction flows $V$ are given by:
\begin{xalignat}{3}
\label{eq:open_V}
  V &= \kkappa \lb V^+_0 - V^-_0 \rb + V^E & \text{where }  V^+_0 &=
  \Exp \Nf^T \Ln K \hp X & 
  V^-_0 &=   \Exp \Nr^T \Ln K \hp X 
\end{xalignat}
where the composite stoichiometric matrices  $N^f$ and $N^r$, the composite
state $X$ and the vector of free-energy constants $K$ are given by:
\begin{xalignat}{4}\label{eq:composite}
  N^f &=
  \begin{pmatrix}
    N_i^f \\ N_e^f
  \end{pmatrix}&
  N^r &=
  \begin{pmatrix}
    N_i^r \\ N_e^r
  \end{pmatrix}&
  X &=
  \begin{pmatrix}
    X \\ X_e
  \end{pmatrix}&
  K &= 
  \begin{pmatrix}
    K_i \\ K_e
  \end{pmatrix}
\end{xalignat}
%
In the context of the simple example shown in Figure \ref{subfig:ABCAo_abg}
it can be verified that substituting the stoichiometric matrices of Equation
\eqref{eq:N_simple} into Equation \eqref{eq:open_V} leads to Equations
\eqref{eq:intro:closed_x} and \eqref{eq:intro:closed_v}.
Equation \eqref{eq:open_V} expresses the forward and reverse reaction
flows $V^+_0$ and  $V^-_0$ in terms of \emph{species}: amounts $X$ and
free-energy constants $K$. It is helpful to reexpress $V^+_0$ and
$V^-_0$ in terms of quantities related to \emph{reactions}.
In particular:
\begin{xalignat}{3}
  V^+_0 &=  K^f \hp X^f &
  \text{where } K^f &= \Exp \Nf^T \Ln K &
  \text{ and } X^f &= \Exp \Nf^T \Ln X \label{eq:KfXf}\\
  V^-_0 &=  K^r \hp X^r &
  \text{where } K^r &= \Exp \Nr^T \Ln K &
  \text{ and } X^r &= \Exp \Nr^T \Ln X \label{eq:KrXr}
\end{xalignat}
In the context of the simple example of Figure \ref{subfig:ABCAo_abg}
\begin{xalignat}{2}
  \label{eq:simple_N_fr}
  \Nf^T &=
  \begin{pmatrix}
    1 & 0 & 0 & 0 & 0\\
    0 & 1 & 0 & 0 & 0\\
    0 & 0 & 1 & 1 &  0
  \end{pmatrix}&
  \Nr^T &=
  \begin{pmatrix}
    0 & 1 & 0 & 0 & 0\\
    0 & 0 & 1 & 0 & 0\\
    1 & 0 & 0 & 0 &  1
  \end{pmatrix}
\end{xalignat}
hence
\begin{xalignat}{4}
  K^f &=
  \begin{pmatrix}
    K_a\\K_b\\K_cK_f
  \end{pmatrix} &
  K^r &=
  \begin{pmatrix}
    K_b\\K_c\\K_aK_r
  \end{pmatrix} &
  X^f &=
  \begin{pmatrix}
    x_a\\x_b\\x_cx_f
  \end{pmatrix} &
  X^r &=
  \begin{pmatrix}
    x_b\\x_c\\x_ax_r
  \end{pmatrix}
\end{xalignat}

\subsection{Energy flow}
\label{sec:energy-flow}
\begin{figure}[htbp]
  \centering
  \SubFig{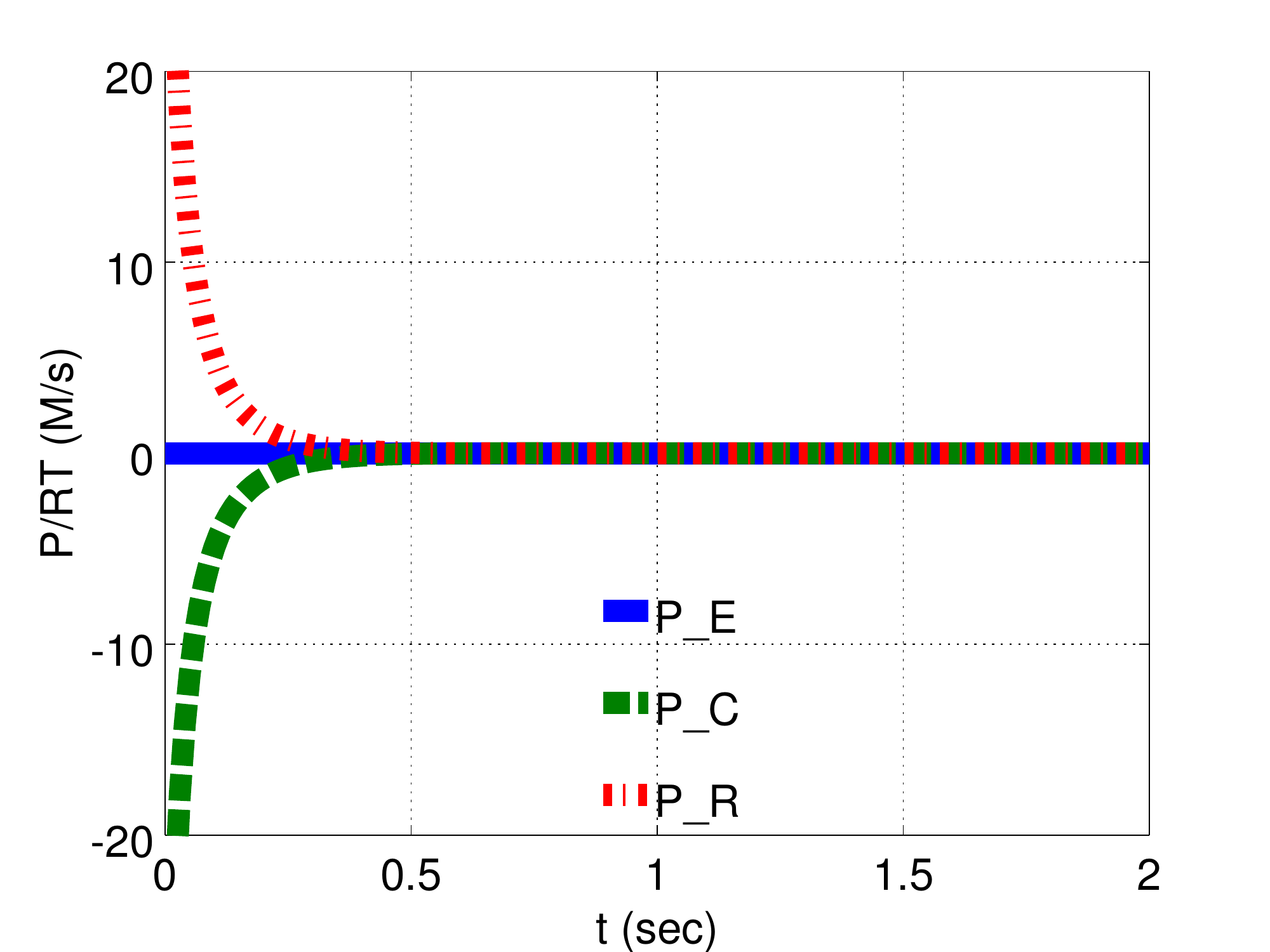}{Closed system}{0.45}
  \SubFig{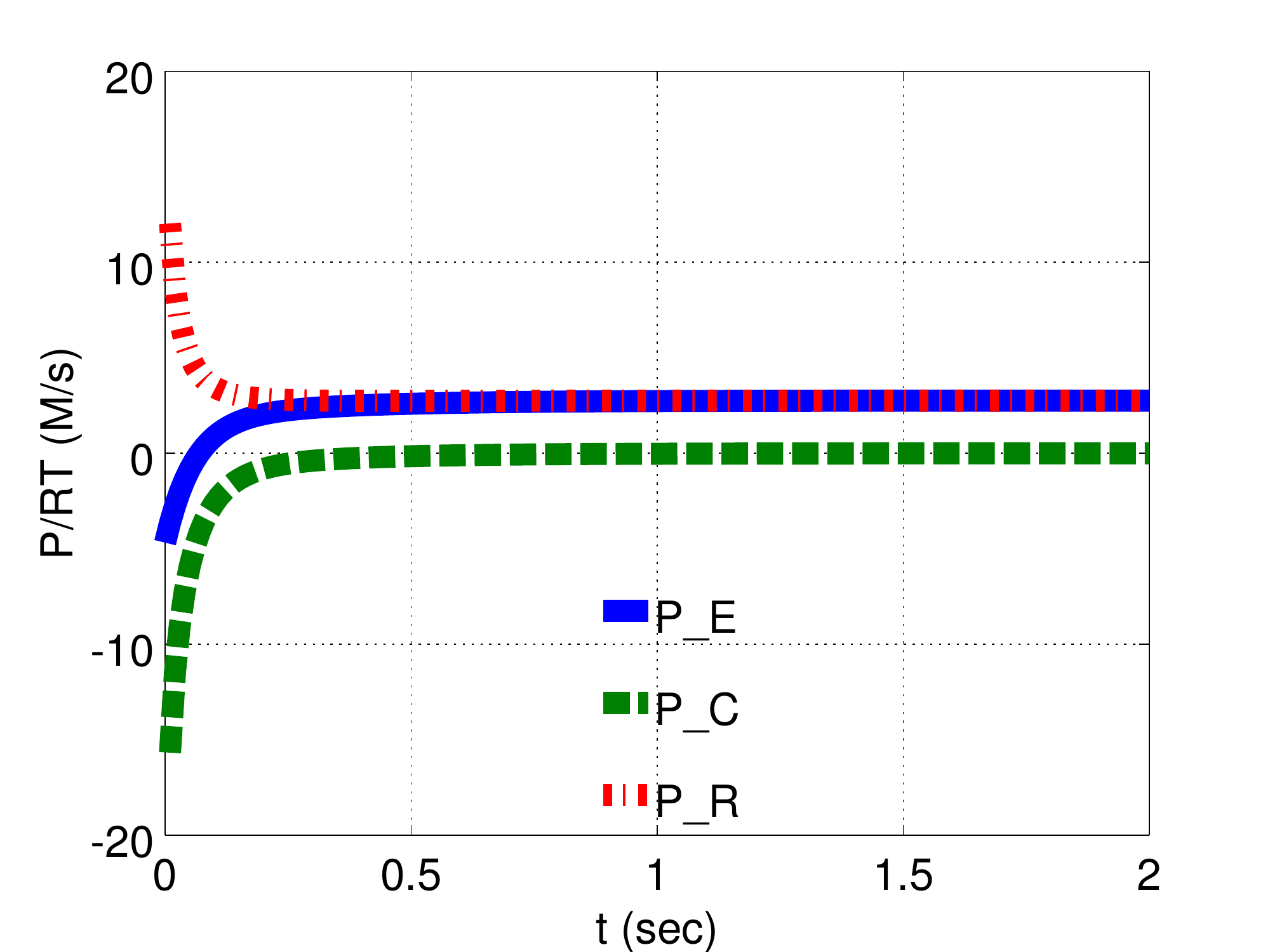}{Open system}{0.45}
  \caption{Energy flows for the system illustrated in Figure \ref{subfig:ABCAo_abg} and simulated in Figure \ref{fig:sim_ABCAo}:
    $P_E$ is the external energy flowing into the system, 
    $P_C$ is the energy flowing into the three \C
    components, and $P_R$ is the energy flowing into (and dissipated by)
    the three \Re components.
    (a) There is no external energy source in the closed system so
    $P_E=0$. Conservation of energy gives $P_C+P_R=0$: the energy
    flowing out of the \C components is dissipated in the \Re components.
    (b) There is an external energy source in the open system with
    $\mu_e = RT \Ln K_e$. Initially, energy flows out of the the open
    system but, in the steady state, the energy into the system is
    balanced by the energy dissipated in the \Re components and the
    energy flow into the \C components becomes zero.  }
  \label{fig:energy}
\end{figure}
With reference to Figure \ref{subfig:Open_abg}: the bonds (\vbond) and
transformers ($\mathcal{TF}$)
transmit, but do not create or destroy, chemical energy; the
$\mathcal{C}$ components store, but do not create or destroy, chemical
energy; the reactions ($\mathcal{R}e$) dissipate, but do not store, chemical
energy; and the ports ($\mathcal{SS}$) transmit chemical
energy in and out of the system.

The energy dissipated within the \Bg{Re} components represented by
$\mathcal{R}e$ is:
\begin{xalignat}{2}
  \label{eq:P_re_0}
  P_R &= \sum_{j=1}^{n_v} \lb v_j A^f_j -  v_j A^r_j \rb = V^T A&
\text{where } A &= A^f-A^r
\end{xalignat}
Using Equations \eqref{eq:affinities}, the affinity $A_i$ 
can be rewritten as:
\begin{equation}
  \label{eq:affinity_i}
  A_i =  A_i^f - A_i^r =  \lb {N_i^f}^T - {N_i^r}^T \rb \mu_i = -N_i^T
  \mu_i = -N_i^T \RT \ln \KK_i X_i
\end{equation}
Similarly and $A_e$
and $A$ can be rewritten as:
\begin{xalignat}{2}
  \label{eq:affinity}
  A_e &= -N_e^T
  \mu_e = -N_e^T \RT \ln \KK_e X_e&
\text{and }   A &= -N^T \mu = -N^T \RT \ln \KK X
\end{xalignat}

Hence naming the external energy flow into the system as $P_E$ and the
energy flow into the \C components as $P_C$ it follows that:
\begin{align}
  P_E &= \RT V^T N_e^T  \Ln K_e \hp X_e \label{eq:P_E}\\
  P_C &= \RT V^T N_i^T \Ln K_i \hp X_i \label{eq:P_C}\\
\text{and }  P_R &= \RT V^T N^T \Ln K \hp X \label{eq:P_R}
\end{align}
These three equations imply energy conservation:
\begin{equation}
  \label{eq:energy-cons}
  P_E = P_C + P_R
\end{equation}
Figure \ref{fig:energy} is based on data corresponding to the
simulations of Figure \ref{fig:sim_ABCAo} and illustrates these
equations for both closed (no external energy source) and open systems.

\section{Conversion of kinetic data}
\label{sec:conv-kinet-data}
For the bond graph  formulation of this paper to be widely applicable,
it is necessary to develop a framework for converting kinetic data from 
the conventional form used with enzyme modelling into the parameters 
required by the bond graph formulation. 

As an example of the issues involved, the reaction $A \reac B$ has a
single equilibrium constant $K_{eq}$, whereas the bond graph
formulation uses the two thermodynamic constants $K_a$ and $K_b$. In
this case, $K_{eq} = \dfrac{K_a}{K_b}$, thus it follows that deriving
$K_a$ and $K_b$ from $K_{eq}$ does not have a \emph{unique} solution.
Similarly, the reaction $A \reac B \reac C \reac A$ of Equation 
\eqref{eq:ABCA_reac} has three equilibrium constants. In this case, 
the solution for $K_a$, $K_b$ and $K_c$ does not \emph{exist} unless 
the product of the three equilibrium constants is unity (detailed balance).
In \S~\ref{sec:equil-const} a derivation of
the formulae that convert equilibrium constants to thermodynamic constants 
for arbitrary networks of reactions is shown. The solution fully 
accounts for the potential of such existence and uniqueness issues.

Enzyme-catalysed reactions are commonly represented using
Michaelis-Menten kinetics \citep{KliLieWie11}.
It should be noted that when a reaction is assumed to be irreversible, the associated
Michaelis-Menten kinetics fail to satisfy thermodynamical compliance. For this 
reason, Michaelis-Menten like kinetics which are thermodynamically compliant have 
been developed \citep{HenBroHat07,EdeGil07,LieUhlKli10}. Section
\ref{sec:conv-kinet-data}\ref{sec:enzyme-catal-react} focuses on the enzyme-catalysed reaction
formulation of \citet[\S5(a)]{GawCra14} and compares it to previously
developed methods.

In the Supplementary Material, this model is shown to be the same as
the as the \emph{direct binding modular} rate law of
\citet{LieUhlKli10} in
\S\ref{app:conv-kinet-data}\ref{sec:relat-direct-bind}, compared with
the \emph{common modular} rate law of \citet{LieUhlKli10} in
\S\ref{app:conv-kinet-data}\ref{sec:relat-comm-modul} and compared
with the the computational model of \citet{LamKus02} in
\S\ref{app:conv-kinet-data}\ref{sec:relat-comp-model}.
As well as the equilibrium constant, mass-action reaction kinetics
also require a rate constant. The corresponding conversion to bond
graph form is discussed in \S
\ref{app:conv-kinet-data}\ref{app:mass-acti-react} of the
Supplementary Material.

\subsection{Equilibrium Constants and Free-energy Constants}
\label{sec:equil-const}
With the notation given in \S~\ref{sec:intro}, the scalar de Donder
formula \citep[Equation(11)]{Bou83} for the ratio $\Vt_0 $ of the
forward $V_0^+$ and backward $V_0^-$ reaction rates of Equation
\eqref{eq:Marcelin} can be written in vector form as:
\begin{xalignat}{2}
   \Vt_0  &= \Exp \frac{A}{\RT} &
   \text{or } \Ln \Vt_0 &= \frac{A}{\RT}\label{eq:deDonder}
\end{xalignat}
where $A = A^f-A^r$
In a similar fashion to \eqref{eq:KfXf} \& \eqref{eq:KrXr} and using
Equation \eqref{eq:affinity}, Equation
\eqref{eq:deDonder} can be written as:
\begin{xalignat}{3}
  \Vt_0 &=  K^{v} \hp X^{v} &
  \text{where } K^{v} &= \Exp \lb {-N}^T \Ln K \rb&
  \text{ and } X^{v} &= \Exp \lb {-N}^T \Ln X \rb \label{eq:KvXv}
\end{xalignat}
Substituting Equation \eqref{eq:KvXv} into Equation
\eqref{eq:deDonder} gives the alternative expression for the affinity $A$:
\begin{equation}
  \label{eq:Affinity_K_v}
  A = \RT \lb \Ln K^{v} \hp X^{v} \rb = \RT \lb \Ln K^{v} + \Ln X^{v} \rb
\end{equation}

In the context of the simple example of Figure \ref{subfig:ABCAo_abg}:
\begin{xalignat}{3}
  {-N}^T &=
  \begin{pmatrix}
    1 & -1 & 0 & 0 & 0 \\
    0 &  1 & -1 & 0 & 0\\
    -1 & 0 &  1 &  1 & -1
  \end{pmatrix}&
  K^{v} &=
  \begin{pmatrix}
    \frac{K_a}{K_b}\\\frac{K_b}{K_c}\\\frac{K_cK_f}{K_aK_r}
  \end{pmatrix} &
  X^{v} &=
  \begin{pmatrix}
    \frac{X_a}{X_b}\\\frac{X_b}{X_c}\\\frac{X_cX_f}{X_aX_r}
  \end{pmatrix} 
\end{xalignat}

As discussed by \citep[Eq 1.15]{KeeSne09}, the equilibrium constant of
a reaction is given by $\exp \frac{\Delta G_0}{RT}$. Combining all
reactions, the vector $K^{eq}$ of equilibrium constants is thus:
\begin{equation}
  \label{eq:Keq_KS}
  K^{eq}  = \Exp \frac{\Delta G_0}{RT} = \Exp \frac{A_0}{RT}
\end{equation}
where the subscript $0$ indicates the reference state where $X$ is
unity.  Hence, substituting unit $X$ into Equation \eqref{eq:KvXv}, Equation \eqref{eq:Keq_KS} becomes:
\begin{equation}
  \label{eq:K_eq}
  K^{eq}  = K^{v} = \Exp \lb {-N}^T \Ln K \rb
\end{equation}
Equilibrium corresponds to $A = 0$ or $\Vt_0=1_{n_v\times1}$ hence, using
Equation \eqref{eq:KvXv}, equilibrium also corresponds to:
\begin{xalignat}{2}
  \label{eq:equilibrium}
  X^{v} &= X^{eq} & \text{where } K^{eq} X^{eq} = 1_{n_v\times1}
\end{xalignat}

Equation \eqref{eq:K_eq} gives an explicit expression for the vector
$K^{eq}$, containing $n_V$ reaction equilibrium constants, in terms
of the vector $K$ which contains the free energy constants of $n_X$ species.

However, the transposed stoichiometric matrix $N^T$ is not normally
full rank, and so it is \emph{not} possible to directly use Equation
\eqref{eq:K_eq} to give $K$ in terms of $K^{eq}$. But, as is now
demonstrated, the left and right null space matrices (as used to
detect conserved moieties and flux pathways \citep{Pal06,Pal11}) lead
to the solutions (if any) of Equation \eqref{eq:K_eq} giving $K$ in
terms of $K^{eq}$.

The $n_v \times n_R$ right null-space matrix $R$ of $N$ has the property that:
\begin{xalignat}{2}
  NR &= 0 &
  \text{or } R^T N^T &= 0 \label{eq:RN}
\end{xalignat}
This matrix is used in stoichiometric analysis to analyse metabolic
pathways \citep{Pal06,Pal11}. Here, it is reused to examine thermodynamic constraints.
Multiplying equation \eqref{eq:K_eq} by $R^T$ gives:
\begin{align}
  R^T \Ln K^{eq} &= 0 \label{eq:constraint}
\end{align}
Equation \eqref{eq:constraint} defines a thermodynamic constraint on
the equilibrium constants, and is a form of
\emph{Wegscheider condition} \citep{LieUhlKli10,KliLieWie11}.

Assuming that the constraint \eqref{eq:constraint} holds, the
Moore-Penrose generalised inverse \citep[\S 6.1]{Ber05}
$N^\dagger$ of $-N^T$ can be used to find a solution for $K$. In
particular:
\begin{xalignat}{2}
 \Ln  K_0 &=  N^\dagger \Ln K^{eq} &
\text{or } K_0 &= \Exp (N^\dagger \Ln K^{eq}) \label{eq:Kc0}
\end{xalignat}

In general, the solution of Equation \eqref{eq:Kc0}, $K=K_0$, is not
the only value of $K$ satisfying Equation \eqref{eq:K_eq}. 
The $n_G \times n_X$ left null-space matrix $G$ (as used to detect
conserved moieties \citep{Pal06,Pal11}) has the property that:
\begin{xalignat}{2}
  GN &= 0 &
  \text{or } N^T G^T &= 0 \label{eq:GN}
\end{xalignat}
Hence a family of solutions of Equation \eqref{eq:K_eq} is given by:
%
\begin{align}
  \Ln K &=    \Ln K_0  +  \Ln K_1 = \Ln K_0{\hp}K_1\label{eq:lnKK}\\
\text{where } \Ln K_1 &= G^T \Ln \kk^c_1\\
\text{or } K_1 &= \Exp ( G^T \Ln \kk^c_1 )\label{eq:lnKK_1}
\end{align}
where $\kk^c_1$ is an arbitrary $n_G \times 1$ vector.
Equation \eqref{eq:lnKK} can be rewritten as:
\begin{align}
  K &= K_0{\hp}K_1 \label{eq:KK_c}
\end{align}

Finally, we show that the affinity $A$ is also unaffected by the
choice of $\kk^c_1$. From Equations \eqref{eq:affinity} and
\eqref{eq:lnKK}--\eqref{eq:KK_c} it follows that:
\begin{align}
  A &= -N^T RT \Ln (K{\hp}X) = RT \Ln K_0{\hp}K_1{\hp}X
  = RT ( -N^T \Ln K_0  -N^T \Ln K_1  -N^T \Ln X )
\end{align}
Using \eqref{eq:GN} \& \eqref{eq:lnKK_1},  $N^T \Ln K_1 = 0$.
Hence $A$ is unaffected by the choice of $\kk^c_1$. It follows that
the energy-based analysis of \S~\ref{sec:energy-flow} is also
unaffected by the choice of $\kk^c_1$.

\paragraph{Example}
The closed system embedded in Figure \ref{subfig:ABCAo_abg} has
stoichiometric matrix $N$, and left and right null space matrices $G$
and $R$ given by:
\begin{xalignat}{3}
  \label{eq:example_NGR}
  N &= N_i =
  \begin{pmatrix}
    -1 & 0 & 1\\
    1 & -1 & 0\\
    0 & 1 & -1
  \end{pmatrix}&
  G &=
  \begin{pmatrix}
    1&1&1
  \end{pmatrix}&
  R &=
  \begin{pmatrix}
    1\\1\\1
  \end{pmatrix}
\end{xalignat}
With this value of $R$, Equation \eqref{eq:constraint} implies that:
\begin{xalignat}{2}
  \label{eq:c_example_constraint}
  \ln K^{eq}_1 + \ln K^{eq}_2 + \ln K^{eq}_1 &= 0 & \text{or }
  K^{eq}_1K^{eq}_3K^{eq}_3 &= 1
\end{xalignat}
This is the standard ``detailed balance'' result applied to a
three-reaction loop \citep[\S 1.3]{KeeSne09}.

Taking the example further, suppose $K^{eq} = (1,\, 0.5,\, 2)^T$ (which satisfies
Equation \eqref{eq:c_example_constraint}). Then using the GNU Octave
implementation \texttt{pinv} to calculate the pseudo inverse for Equation
\eqref{eq:Kc0} gives $K_0 = (0.79370,\,0.79370,\,1.58740)^T$. Noting that
$n_G=1$, $k_1$ is a scalar and may be chosen as $k_1 = 1/0.79370$ leading
to $K = (1,\,1,\,2)^T$.

In contrast, the open system of Figure \ref{subfig:ABCAo_abg} has
stoichiometric matrix $N$, and left and right null space matrices $G$
and $R$ are given by
\begin{xalignat}{3}
  \label{eq:example_NGR_open}
  N &=
  \begin{pmatrix}
    -1 & 0 & 1\\
    1 & -1 & 0\\
    0 & 1 & -1\\
    0 & 0 & -1\\
    0 & 0 & 1\\
  \end{pmatrix}&
  G &=
  \begin{pmatrix}
    1&1&1&0&0\\
    0&0&0&1&1
  \end{pmatrix}&
  R &= 0
\end{xalignat}
Because $R=0$, the constraint \eqref{eq:constraint} holds for any
choice of the three elements of $K^{eq}$. For example, suppose $K^{eq} = (1,\, 0.5,\, 16)^T$,
then using the GNU Octave
implementation \texttt{pinv} for the pseudo inverse, Equation
\eqref{eq:Kc0} gives $K_0 = (0.79370,\,0.79370,\,   1.58740,\,   2.82843,\,   0.35355)^T$.
Noting that
$n_G=2$, $k_1$ has two elements and may be chosen, for example, as $k_1 = (1/0.79370,\,2/1.58740$ leading
to $K = (1,\, 1,\, 2,\, 2,\, 0.25)^T$.


\subsection{Enzyme Catalysed Reactions}
\label{sec:enzyme-catal-react}
The kinetics of enzyme catalysed reactions are usually described by
versions of the Michaelis-Menten equations; see, for example,
\citet[\S2.1]{KliLieWie11} or
\citet[\S1.4]{KeeSne09}. 
The key issue is that the equations should be
thermodynamically compliant \citep{LieUhlKli10}.

\citet[\S5(a)]{GawCra14} give one possible formulation of the kinetics for enzyme catalysed
reactions that is guaranteed to be thermodynamically compliant:
\begin{xalignat}{2} 
  v &= \bar{\kappa} \frac{K_ee_0}{1 + \frac{\sigma_v}{k_v}}
 \delta_v &
\text{ where }
 \sigma_v = \frac{\kappa_1 V_0^+ + \kappa_2 V_0^-}{\kappa_1  +
   \kappa_2}\label{eq:v_GC}
\end{xalignat}
where $V_0^+$ and $V_0^-$ are given by Equation \eqref{eq:Marcelin}, 
$e_0$ is the total enzyme concentration, and $\kappa_1$ and $\kappa_2$
are reaction constants for the reactions relating substrate to complex
and complex to product respectively. $k_v$ is a constant related to
the enzyme/complex equilibrium.  As is now shown this formulation can
be rewritten as a reversible Michaelis-Menten equation.
Defining:
\begin{equation}
  \rho_v =
\frac{\kappa_2}{\kappa_1 + \kappa_2} \text{ gives }
\sigma_v = (1-\rho_v)  V_0^+  + \rho_v V_0^-
\end{equation}
With reference to Equation \eqref{eq:v_GC}, define:
\begin{equation}
  \label{eq:v_max}
  v_{max}^+ = \lim_{V_0^+ \rightarrow \infty} v =
  \frac{v_{max}}{1-\rho_v}\;
  \text{ and }   v_{max}^- = \lim_{V_0^- \rightarrow \infty} v =
  \frac{v_{max}}{\rho_v}\; 
  \text{ where } v_{max} = \bar{\kappa}k_ce_0
\end{equation}
then Equation \eqref{eq:v_GC} can be rewritten as:
\begin{equation}
  \label{eq:v_MM_0}
  v = \frac
  {v_{max}^+\frac{V_0^+}{k_v^+} - v_{max}^-\frac{V_0^-}{k_v^-}}
  {1 + \frac{V_0^+}{k_v^+} + \frac{V_0^-}{k_v^-}}
\text{ where }
k_v^+ = \frac{k_v}{(1-\rho_v)}\;,
k_v^- = \frac{k_v}{\rho_v}
\end{equation}
Combining Equations \eqref{eq:v_max} and \eqref{eq:v_MM_0}:
\begin{equation}
  \label{eq:v_MM}
  v = \frac{v_{max}}{k_v} \frac{V_0^+ - V_0^-}{1 + \frac{1}{k_v}
    \lb \lb 1 - \rho_v \rb V_0^+ + \rho_v  V_0^-\rb}
= \frac{v_{max}  \lb V_0^+ - V_0^- \rb}{k_v + 
    \lb 1 - \rho_v \rb V_0^+ + \rho_v  V_0^-}
\end{equation}

\paragraph{Remarks}
\begin{enumerate}
\item Equation \eqref{eq:v_MM} explicitly shows that this particular
  form of the reversible Michaelis-Menten equation has three
  parameters: $v_{max}$, $k_{v}$ and $\rho_v$.
\item When $\rho_v = 0$, Equation \eqref{eq:v_MM} becomes the
  irreversible Michaelis-Menten equation but with the addition of the
  $V_0^-$ term to give reversibility.
\item Using Equation \eqref{eq:v_max}, the two parameters $v_{max}$
  and $\rho_v$ can be computed from $v_{max}^+$ and $v_{max}^-$ as:
  \begin{equation}
    \label{eq:v_max_rho}
    \rho_v = \frac{\frac{v_{max}^+}{v_{max}^-}}{1+\frac{v_{max}^+}{v_{max}^-}} =
    \frac{v_{max}^+}{v_{max}^+ + v_{max}^-}
\text{ and } v_{max} = \frac{v_{max}^+v_{max}^-}{v_{max}^+ + v_{max}^-}
  \end{equation}
\item Using Equation \eqref{eq:v_MM_0}, the third parameter $k_v$ can
  be computed from:
  \begin{equation}
    \label{eq:k_v}
    k_v = \lb 1 - \rho_v \rb k_v^+ = \rho_v k_v^-
  \end{equation}
\item Note that the Haldane equation
  $K_{eq}=\dfrac{v_{max}^+k_v^-}{v_{max}^-k_v^+}$ is automatically
  satisfied by Equation \eqref{eq:v_MM}.
\item In the limit  as $k_v \rightarrow \infty$, Equation
  \eqref{eq:v_MM} becomes the mass-action flow \citep[Equation (2.6) ]{GawCra14}:
  \begin{equation}
    \label{eq:MM-MA}
    v = \kappa \lb V_0^+ - V_0^- \rb
    \text{ where } \kappa = \frac{v_{max}}{k_v} 
  \end{equation}
\end{enumerate}

\section{Hierarchical modelling}
\label{sec:hier-modell}
\begin{figure}[htbp]
  \centering
  \Fig{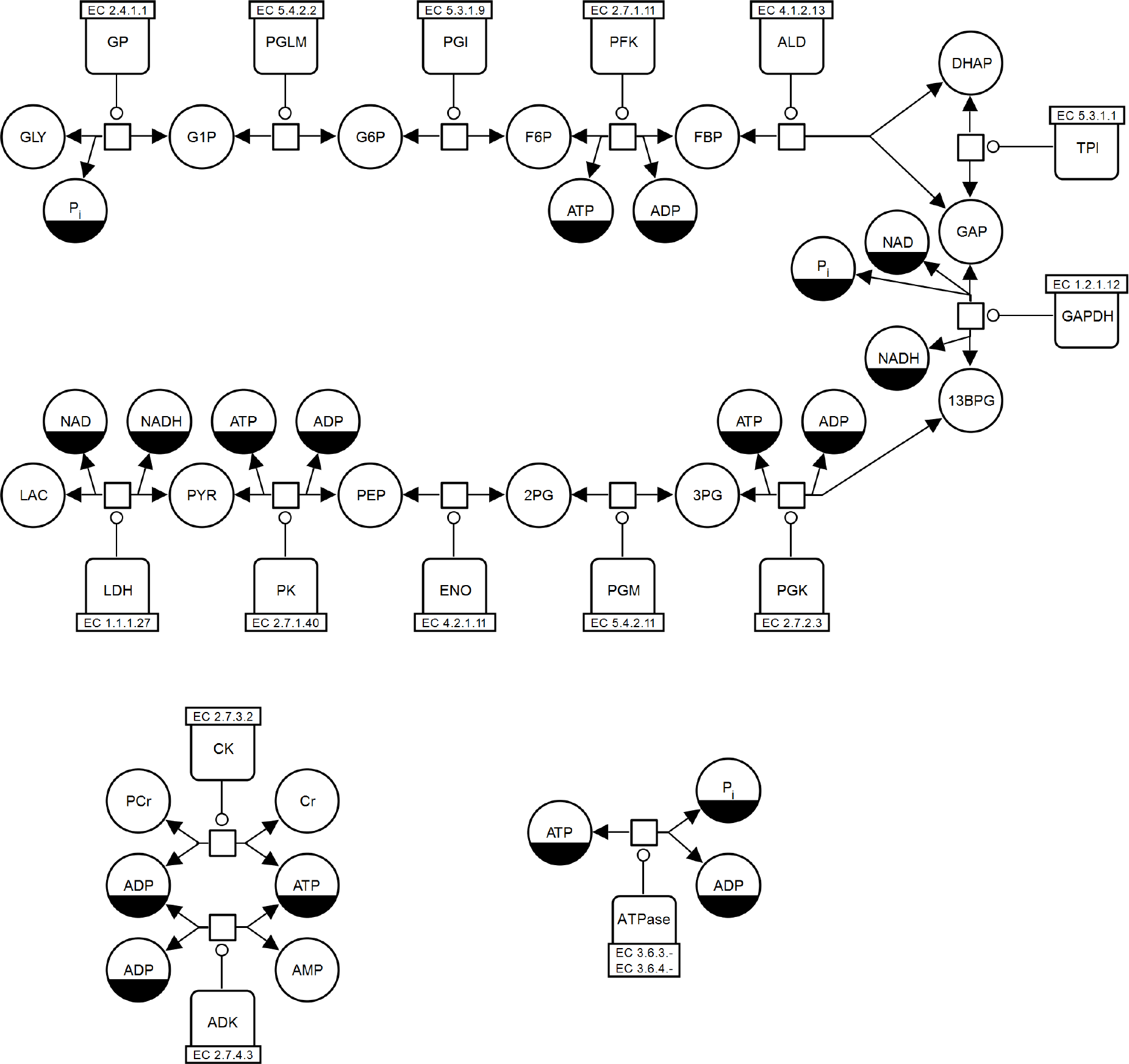}{0.9}
\caption{The simplified glycolytic pathway as described by \citet[Figure 1]{LamKus02}. 
  \textbf{Reactions/enzymes:} GP, glycogen phosphorylase; PGLM, phosphoglucomutase; PGI, 
  phosphoglucose isomerase; PFK, phosphofructo kinase; ALD, aldolase; TPI, 
  triose phosphate isomerase; GAPDH, glyceraldehyde 3-phosphate dehydrogenase;
  PGK, phosphoglycerate kinase; PGM, phosphoglycerate mutase; ENO, enolase; 
  PK, pyruvate kinase; LDH, lactate dehydrogenase; CK, creatine kinase; ADK, 
  adenylate kinase; ATPase, ATPases. \textbf{Metabolites:} GLY, glycogen;
  P$_{i}$, inorganic phosphate; G1P, glucose-1-phosphate; G6P, glucose-6-phosphate;
  F6P, fructose-6-phosphate; ATP, adenosine triphosphate; ADP, adenosine diphosphate;
  FBP, fructose-1,6-biphosphate; DHAP, dihydroxyacetone phosphate; GAP, 
  glyceraldehyde 3-phosphate; NAD, oxidised nicotinamide adenine dinucleotide; NADH, 
  reduced NAD; 13BPG, 1,3-bisphosphoglycerate; 3PG, 3-phosphoglycerate; 2PG, 
  2-phosphoglycerate; PEP, phosphoenolpyruvic acid; PYR, pyruvate; LAC, lactate;
  PCr, phosphocreatine; Cr, creatine; AMP, adenosine monophosphate. Enzyme 
  commission (EC) numbers are shown.}
  \label{fig:LamKus02}
\end{figure}
\begin{figure}
  \centering
  \SubFig{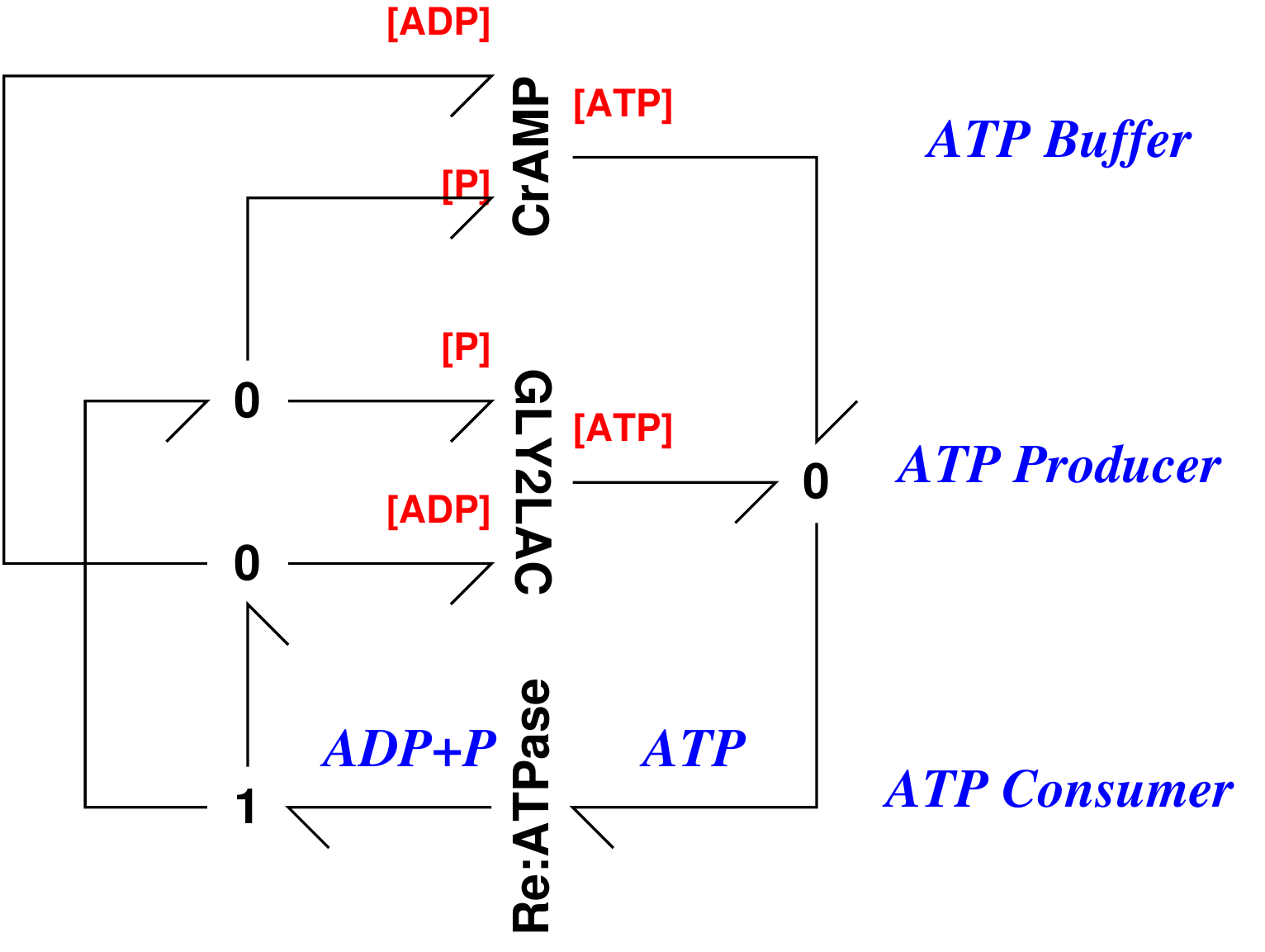}{\textbf{LamKus02}}{0.45}
  \SubFig{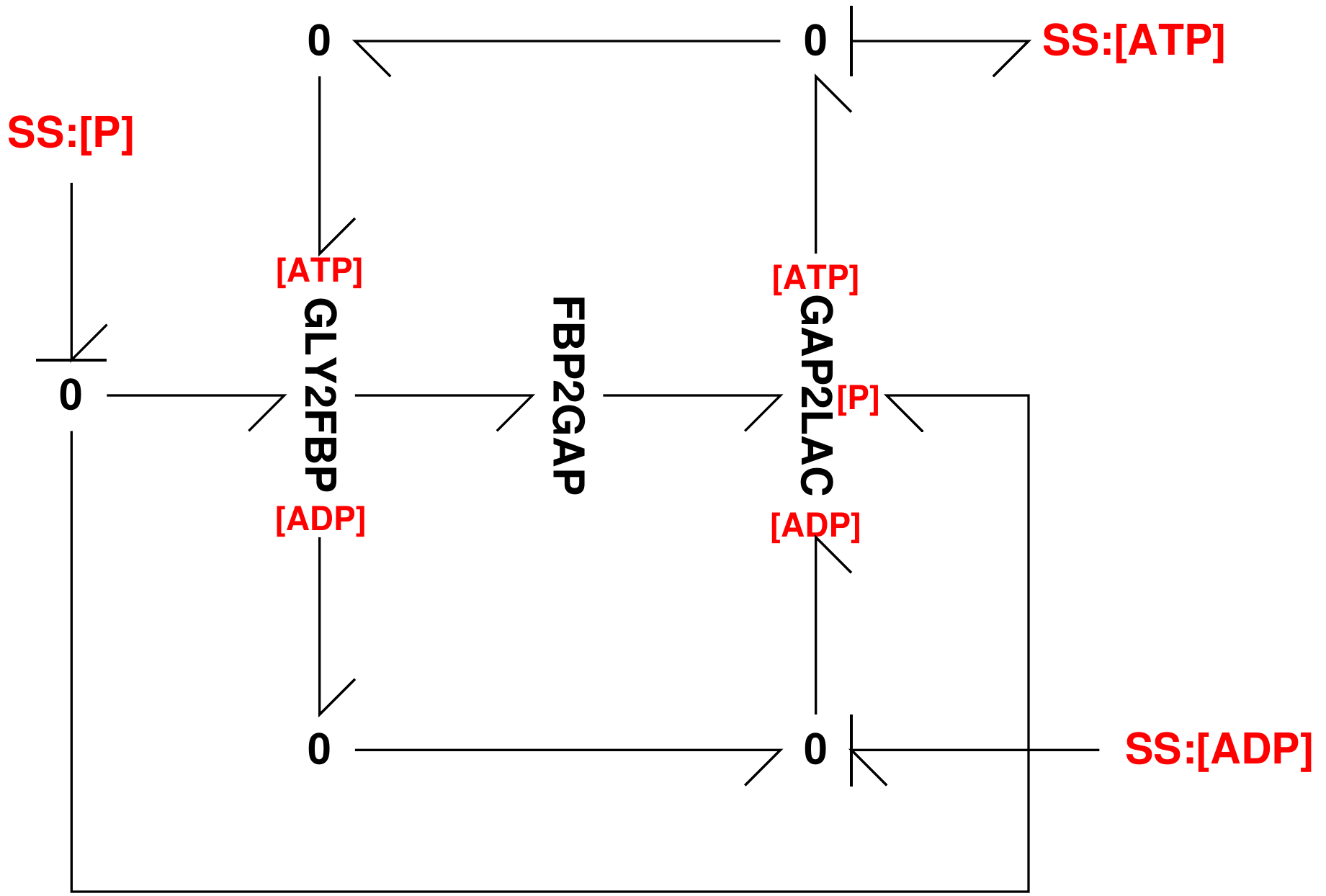}{\textbf{GLY2LAC}}{0.45}
  \SubFig{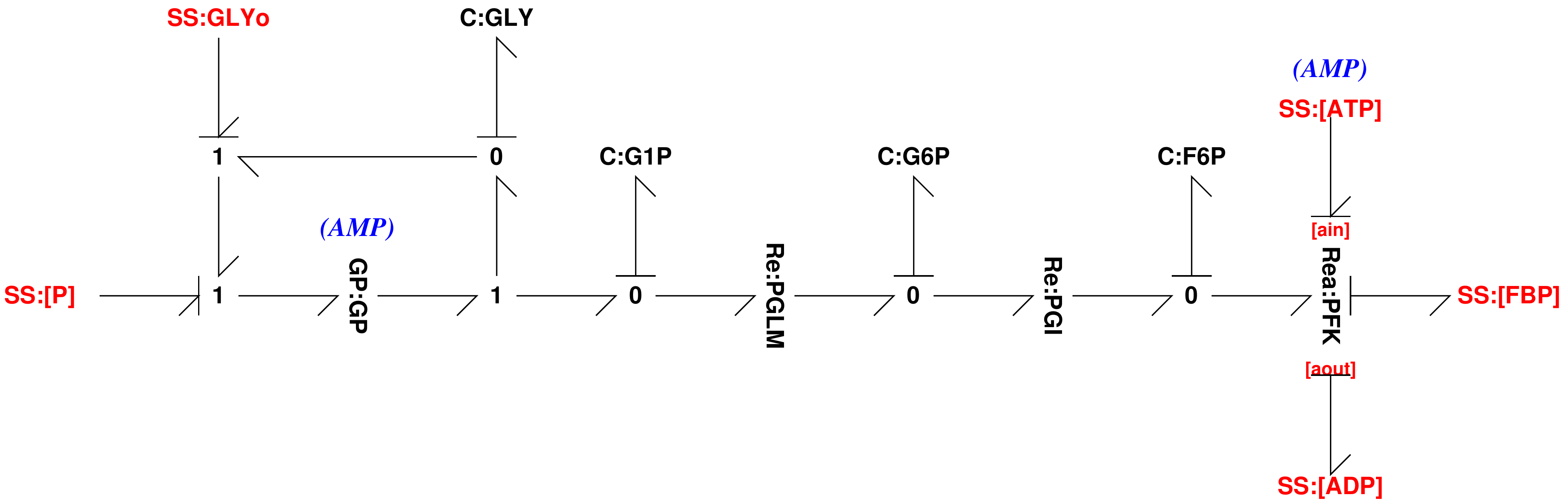}{\textbf{GLY2FBP}}{0.9}
  \SubFig{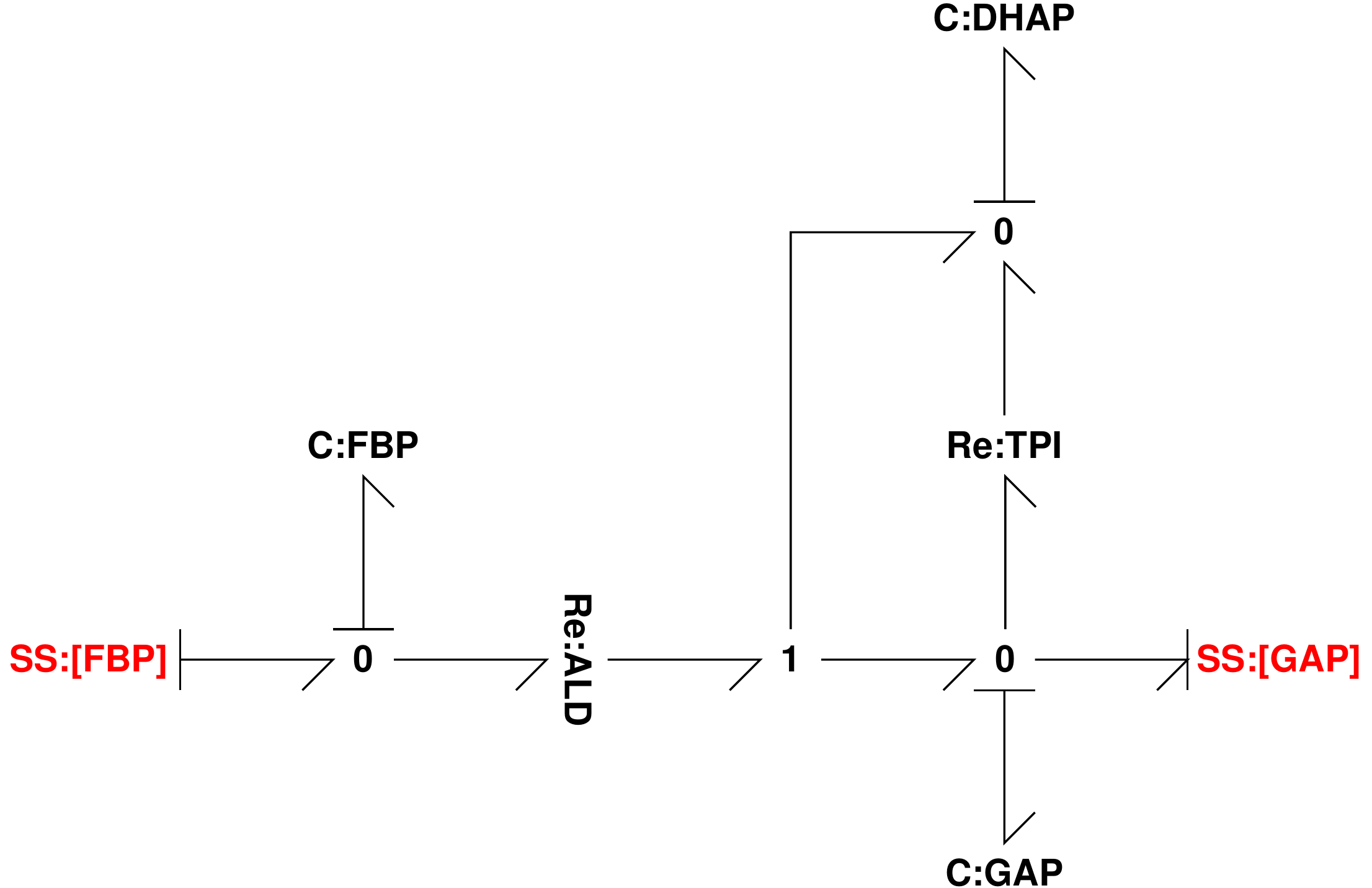}{\textbf{FBP2GAP}}{0.45}
  \SubFig{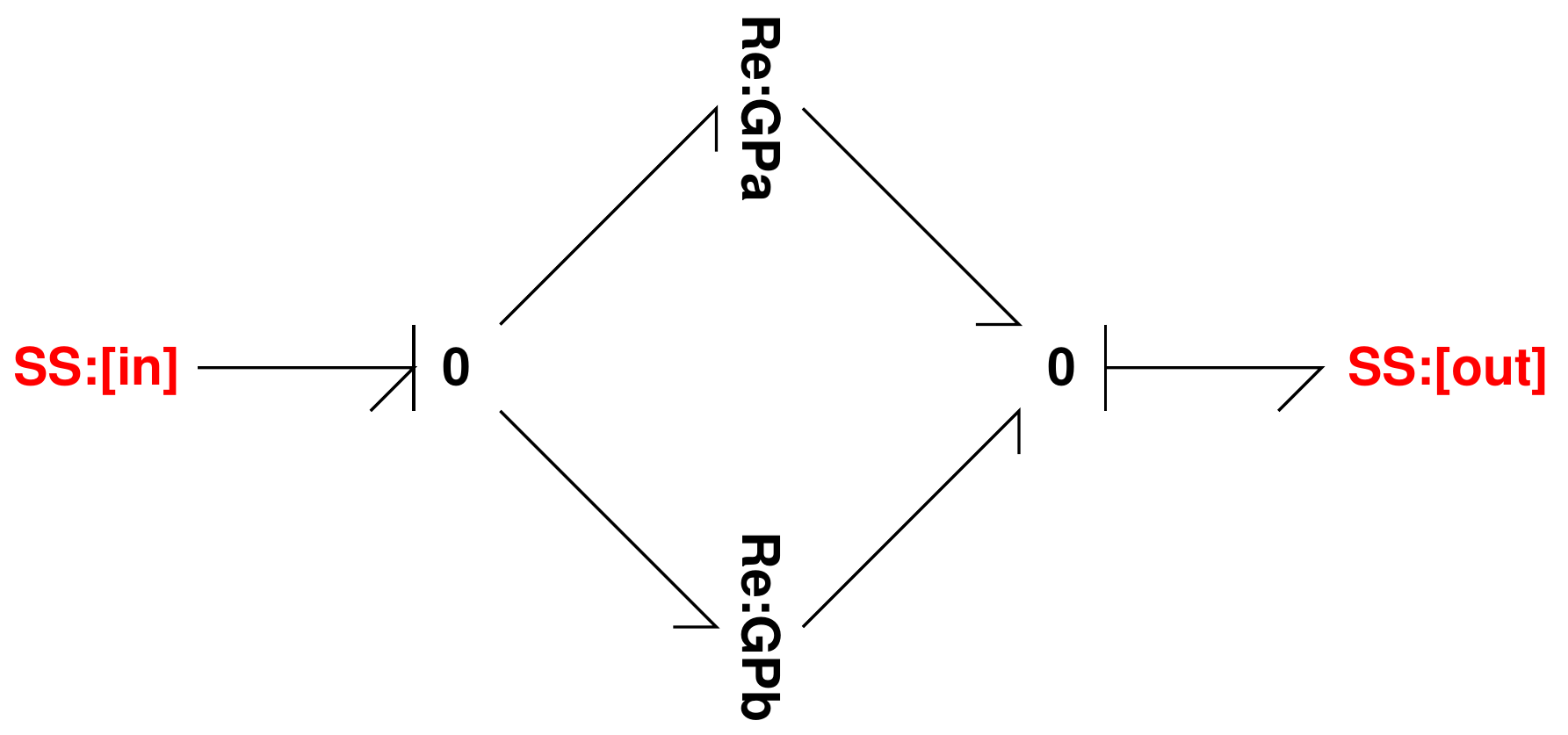}{\textbf{GP}}{0.45}
  \caption{Hierarchical Bond Graph Model. (a) As discussed in the
    text, \textbf{LamKus02} represents the top-level model of the
    system in Figure \ref{fig:LamKus02}. (b) \textbf{GLY2LAC}
    represents one of the three submodels in (a). (c)\&(d)
    \textbf{GLY2FBP}\&\textbf{FBP2GAP} represent two of the three
    submodels in (b). (e) The reaction $GP$ has two parallel reactions
    $GPa$ and $GPb$.}
  \label{fig:BG}
\end{figure}


The bond graph representation for open systems detailed in \S
\ref{sec:closed-open-systems} Figure \ref{fig:closed-open-systems},
with the $\mathcal{SS}$ ports to connect systems, provides a basis to
construct hierarchical models of biochemical systems that are robustly
thermodynamically compliant. This approach is illustrated using a
well-established model from the literature: ``A Computational Model
for Glycogenolysis in Skeletal Muscle'' presented by \citet{LamKus02}.
Although the model has been further embellished by \citet{VinRusPal10}
and used as an example in the book of \citet{Bea12}, we use
information and parameters from the original model as a basis for the
discussion in this paper.

Figure \ref{fig:LamKus02} shows the simplified glycolysis pathway
from \citet[Figure 1]{LamKus02} using Systems Biology Graphical
Notation (SBGN) \citep{NovHukMi09}.
There are many ways to subdivide this system to create a hierarchical
model.  Here, we have chosen to divide the system into three
conceptual
modules:
\begin{enumerate}
\item The primary glycolytic reaction chain leading from glycogen 
  ($GLY$) to lactate ($LAC$) which converts adenosine diphosphate
  ($ADP$) and inorganic phosphate ($P$) into adenosine triphosphate
  ($ATP$) making use of the energy stored in glycogen. This module is
  a \emph{producer} of ATP.
\item The pair of reactions catalysed by creatine kinase $CK$ and
  adenylate kinase $ADK$ involving creatine ($Cr$), phosphocreatine
  ($PCr$) and adenosine monophosphate ($AMP$) as well as $ATP$ and
  $ADP$. This module is a \emph{buffer} of ATP.
\item The reactions catalysed by numerous ATPases ($ATPase$; for a more
  comprehensive description see \citet{LamKus02}) which convert
   $ATP$ into $ADP$ and $P$, and use the released energy to perform work. This module is
  a \emph{consumer} of ATP.
\end{enumerate}
Figure \ref{fig:BG} gives a bond graph representation of this
top-level decomposition where the three modules are represented by the
compound bond graph components labelled \textbf{GLY2LAC},
\textbf{CrAMP} and the simple reaction bond graph component
\textbf{Re:ATPase}. The metabolites $ATP$, $ADP$ and $P$ flow between
these three modules as illustrated, forming the overall system model
\textbf{LamKus02}\footnote{As discussed by \citet{GawCra14}, junctions
  (such as those appearing in Figure \ref{subfig:GLY2LAC_abg}) with
  only two impinging bonds could be deleted. They are often left in
  place for clarity or to make further connections if the model is
  further refined.}.
The module \textbf{GLY2LAC} is the most complex of Figure
\ref{fig:BG}, and for this reason it is itself hierarchically
decomposed in to three further modules (Figure
\ref{subfig:GLY2LAC_abg}), represented by the compound bond graph
components: \textbf{GLY2FBP}, \textbf{FBP2GAP} and
\textbf{GAP2LAC}. Bond graph representations for these reactions and
species are given in Figure \ref{fig:BG}. The additional component
\textbf{SS:GLYo} is discussed in
\S~\ref{sec:modelling-issues}.
The module corresponding to \textbf{CrAMP} is simpler and is shown
using the bond graph representation of reactions and species in the
Supplementary Material, Figure \ref{fig:CrAMP}.

There are two approximations made in our implementation of the model
described by \citet{LamKus02}: 
\begin{enumerate}
\item the allosteric modulation of reactions $GP$ and $PFK$ by \textbf{AMP} is
  ignored.
\item the reaction kinetics are represented by Equation
  \eqref{eq:v_MM}. As discussed in \S \ref{app:conv-kinet-data}, the
  kinetics presented by \citet{LamKus02} are essentially the common
  modular rate law of \citet{LieUhlKli10} and, as mentioned in
  \S~\ref{sec:enzyme-catal-react}, Equation
  \eqref{eq:v_MM} is essentially the direct binding modular rate law of
  \citet{LieUhlKli10}, our kinetics differ except for first-order
  reactions.
\end{enumerate}
Neither assumption conflicts with our aim of illustrating the creation
of robustly thermodynamically compliant model. As discussed in
\S~\ref{sec:conclusion}, future work will look at the more complicated
case.


\subsection{Modelling issues}
\label{sec:modelling-issues}
  The bond graph approach to modelling imposes discipline on the
  modelling process and thus exposes errors and inconsistencies that
  may otherwise have escaped attention. The process of revising the
  well-established model of \citet{LamKus02}, and its reuse by other
  authors illustrate this process. In particular, four 
  issues were encountered during the development of these bond graph
  models and deserve special attention: the parallel reactions $GPa$
  and $GPb$ forming the overall $GP$ reaction; the modelling of the
  conversion of glycogen $GLY$ to glucose-1-P $G1P$; the reaction
  catalysed by $ATPase$; and reaction directionality.

\paragraph {The parallel reactions $GPa$ and $GPb$}
The parallel reactions $GPa$ and $GPb$ are represented in bond graph
terms in Figure \ref{subfig:GP_abg}. Applying the analysis of \S
\ref{sec:conv-kinet-data}~\ref{sec:equil-const}
to the $22\times17$ stoichiometric matrix $N$ corresponding the entire
network gives a $17\times 1$ right null-space matrix $R$ with an entry
of $-1$ corresponding to $GPa$ and $1$ corresponding to $GPb$. Thus the
condition $R^T \Ln K^{eq} = 0$ of Equation \eqref{eq:constraint} gives:
\begin{xalignat}{2}\label{eq:GP_Weg}
  -\ln K^{eq}_{GPa} + \ln K^{eq}_{GPb} &=0 & \text{or } K^{eq}_{GPa}
  &= K^{eq}_{GPb}
\end{xalignat}
Equation \eqref{eq:GP_Weg} must be satisfied for thermodynamical
compliance, corresponding to the notion that these alternative forms
of the enzyme are catalysing the same chemical reaction. In fact, 
\citet[Table 1]{LamKus02} have $K^{eq}_{GPa}=K^{eq}_{GPb}=0.42$ such 
that condition \eqref{eq:GP_Weg} is satisfied and it is therefore 
possible to convert equilibrium constants of this model to
the free-energy constants required for bond graph modelling.

It should be noted, however, that in its original form the model is 
\emph{not} robustly thermodynamically compliant: if equality \eqref{eq:GP_Weg} 
was violated due to computing or human error, the compliance would not be
enforced. In this context, it is interesting to note that such an
error exists in the published literature, with the paper of
\citet{MosAlfMaj12} re-using the model of \citet{LamKus02}. 
In particular, although on p. 17 \citet{MosAlfMaj12} correctly
use $ K^{eq}_{GPa}=0.42$, the value on p. 18 incorrectly gives
$K^{eq}_{GPb} = 16.62$ thus their model is not thermodynamically
compliant. A glance at \citep[Table 1]{LamKus02} reveals that
\citet{MosAlfMaj12} have inadvertently copied the equilibrium constant
for Phosphoglucomutase instead of that for Glycogen Phosphorylase B.
In contrast, a model expressed in bond graph  form could have
incorrect parameters; but it would still be thermodynamically
compliant. This is the advantage of \emph{robust} thermodynamical
compliance.
\paragraph{The conversion of glycogen ($GLY$) to glucose-1-P ($G1P$)}
\citet{LamKus02} discuss ``uncertainty in the kinetic function and
substrate concentration describing the glycogen phosphorylase
reaction'' and embed a simplified model of the reaction within the
overall model. However (as they explicitly state) the resulting model
is stoichiometrically and thermodynamically inconsistent. Figure
\ref{fig:LamKus02}, and the underlying Figure 1 of \citet{LamKus02}
imply a reaction:
\begin{equation}
  \label{eq:GP_v1}
  GLY + P \reac G1P
\end{equation}
and this is consistent with their equation (p. 821) for the rate of
change of $GLY$: $GLY^\prime = -\text{flux}_{GP}$. However, their
equation (p. 822) for $\text{flux}_{GP} = V_{GP}$ implies the reaction:
\begin{equation}
  \label{eq:GP_v2}
  GLY + P \reac GLY + G1P
\end{equation}
as they explain, this arises by equating two versions of the glycogen molecule
which differ in length by the presence of a single monomer: 
$\text{Glycogen}_n$ and $\text{Glycogen}_{n-1}$. Biologically, this 
reflects the large, polymeric nature of glycogen such that a monomer/subunit 
can be cleaved with little discernible effect. It is explicitly stated as 
an assumption by \cite{LamKus02} that $\text{Glycogen}_n$/$\text{Glycogen}_{n-1}$ 
is unity. Unlike reaction \eqref{eq:GP_v1}, reaction \eqref{eq:GP_v2} implies a 
zero rate of change of $GLY$: $GLY^\prime = 0$.

Using the bond graph approach it is not possible to simultaneously
implement the rate change of $GLY$ corresponding to reaction
\eqref{eq:GP_v1} with the reaction flux implied by
\eqref{eq:GP_v2}. In short, this is because the stoichiometric matrix
associated with reactions \eqref{eq:GP_v1} and \eqref{eq:GP_v2} are
different and thus the assumption of \S
\ref{sec:closed-open-systems}~\ref{sec:general-case} that bonds and
junctions transmit, but do not create or destroy, chemical energy
would be violated. Conceptually, this is equivalent to mass creation, as
glycogen is cleaved by glycogen phosphorylase but does not change. In 
practice the effects would be limited over short simulation times, but 
with a longer term goal of building reusable modular models, it will
ultimately lead to models lacking thermodynamic compliance.

This issue is addressed by introducing the external substance $GLYo$, 
representing the difference between $\text{Glycogen}_n$ and 
$\text{Glycogen}_{n-1}$, into the $GP$ catalysed reactions
\eqref{eq:GP_v1} and \eqref{eq:GP_v2} of \citet{LamKus02} using the bond 
graph component \textbf{SS:GLYo} (Figure \ref{subfig:GLY2FBP_abg}):
\begin{equation}
  \label{eq:GP_bg}
  GLY + GLYo + P \reac GLY + G1P
\end{equation}
Using the notation of \citet{LamKus02}; reaction \eqref{eq:GP_bg} implies 
that $GLYo^\prime = -\text{flux}_{GP}$ and $GLY^\prime = 0$, thus combining 
the two incompatible expressions for $GLY^\prime$ in a stoichiometrically and
thermodynamically consistent way. This illustrates how the bond graph  methodology
forces the modeller to tackle such issues by not allowing a stoichiometrically and
thermodynamically inconsistent model to be constructed.
  Thus this revision replaces a thermodynamically inconsistent
  submodel by a thermodynamically consistent, and therefore physically
  plausible, submodel. This physically plausible submodel acts as a
  placeholder until a more complex chemically-correct submodel can be
  devised. This illustrates the role of the bond graph framework to
  allow incremental refinement of individual submodels within a
  hierarchical system.
\paragraph{The $ATPase$ catalysed reaction}
\citet{LamKus02} state that ``The model design features stoichiometric
constraints, mass balance, and fully reversible thermodynamics as
defined by the Haldane relation.''  However they do break this feature
by choosing the reaction catalysed by $ATPase$ to be irreversible.
Unlike \citet{LamKus02}, the reaction catalysed by $ATPase$
(represented by \textbf{Re:ATPase} in Figure \ref{fig:BG}) is
\emph{reversible} and corresponds to:
\begin{equation}
   ATP \reac  ADP+P 
\end{equation}
Thus the model in this paper is fully thermodynamically reversible and
a closed system.
  Of course this reaction is, for practical purposes, one
  way. But, to follow the systematic bond graph modelling procedure,
  this fact should be reflected by the choice of reaction parameters
  rather than by violating thermodynamic principles. This point is
  discussed in detail by \citet{CorCar00} who state that ``Our view is
  that it is always best to use reversible equations in metabolic
  simulations for all processes apart from exit fluxes ...''.

However, as will be discussed in \S~\ref{sec:sim}, this closed
system may be converted into an open system by injecting external
flows of: $GLYf$, $GLYr$ and $LAC$ in such a way as to make the 
concentrations of these three substances constant. 

The simulation is also arranged such that that flows can be optionally
disconnected to examine system connectivity. This is used in \S
\ref{sec:hier-modell}\ref{sec:sim} to disconnect the reactions $Fout$ 
and $ATPase$ as appropriate.
\paragraph{Reaction directionality}
The $TPI$ reaction within \citet{LamKus02} provides an interesting
example about definitions of reaction ``direction'', which must be
specified within an associated model description. On page 823,
\citet{LamKus02} say that ``The forward direction of $TPI$ is defined
as producing dihydroxyacetone phosphate''. This is why the directions
of the bonds impinging on \textbf{Re:TPI} in Figure
\ref{subfig:FBP2GAP_abg} are in the directions shown. 
\citet{LamKus02} also state that the equilibrium constant for $TPI$ is
$K_{eq} = 0.052$. However, this is inconsistent with the stated
directionality and should be replaced by the reciprocal value (see Supplementary Material,
\S~\ref{app:TPI} for more details).  The
discipline imposed by the bond graph model avoids ambiguity with
regard to reaction direction.

\subsection{Stoichiometric Analysis}
\label{sec:stoich-analys}
\begin{table}[htbp]
  \centering
  \begin{tabular}{|l|l|}
    \hline
    1	&ADP AMP ATP \\
    2	&Cr PCr \\
    3	&NAD NADH \\
    4	&DPG NAD P2G P3G PEP PYR \\
    5	&ADP 2ATP P PCr DHAP 2FBP GAP 2DPG P2G P3G PEP F6P G1P G6P \\
    6	&GLY \\
    \hline
  \end{tabular}
  \caption{Conserved Moieties. The left null matrix of the stoichiometric
    matrix for this system reveals six conserved moieties. The
    simulation takes account of these using 
    Equation \eqref{eq:reduced} in \S~\ref{app:reduc-order-equat}.}
  \label{tab:CM}
\end{table}

As discussed in the  textbooks of 
\citet{Pal06,Pal11} and in the bond graph  context by \citet{GawCra14}
the left null matrix of the system stoichiometric matrix gives in
formation about conserved moieties.
With reference to Table \ref{tab:CM} the bond graph  model has six
conserved moities:
\begin{itemize}
\item CM 1--3 are obvious from the equations and pooled metabolite
  components are assumed to be constant by \citet{LamKus02}.
\item CM 4
  corresponds to the ``total oxidised'' part of
  \citep[Eqn. (3)]{LamKus02}.
\item CM 5 corresponds to
  \citep[Eqn. (2)]{LamKus02} of which they say ``Correct conservation of
  mass within the model was proven for both open and closed systems by
  calculating the total [free] phosphate [note the absence of AMP] using 
  the following equation''.
\item CM 6 arises as the net flow into $GLY$ is zero and reflects the 
  assumption by \citet{LamKus02} that $\text{Glycogen}_n$/$\text{Glycogen}_{n-1}$ 
  is unity.
\end{itemize}
Using the reduced order equations (Supplementary Material, \eqref{eq:reduced}
\S~\ref{app:reduc-order-equat}), these six CMs are
automatically taken into account and even numerical errors cannot
cause drift in these CMs. Reduced order equations are utilised in the
simulations presented in the following \S.

\subsection{Simulation}
\label{sec:sim}
\begin{figure}[htbp]
  \centering
  \SubFig{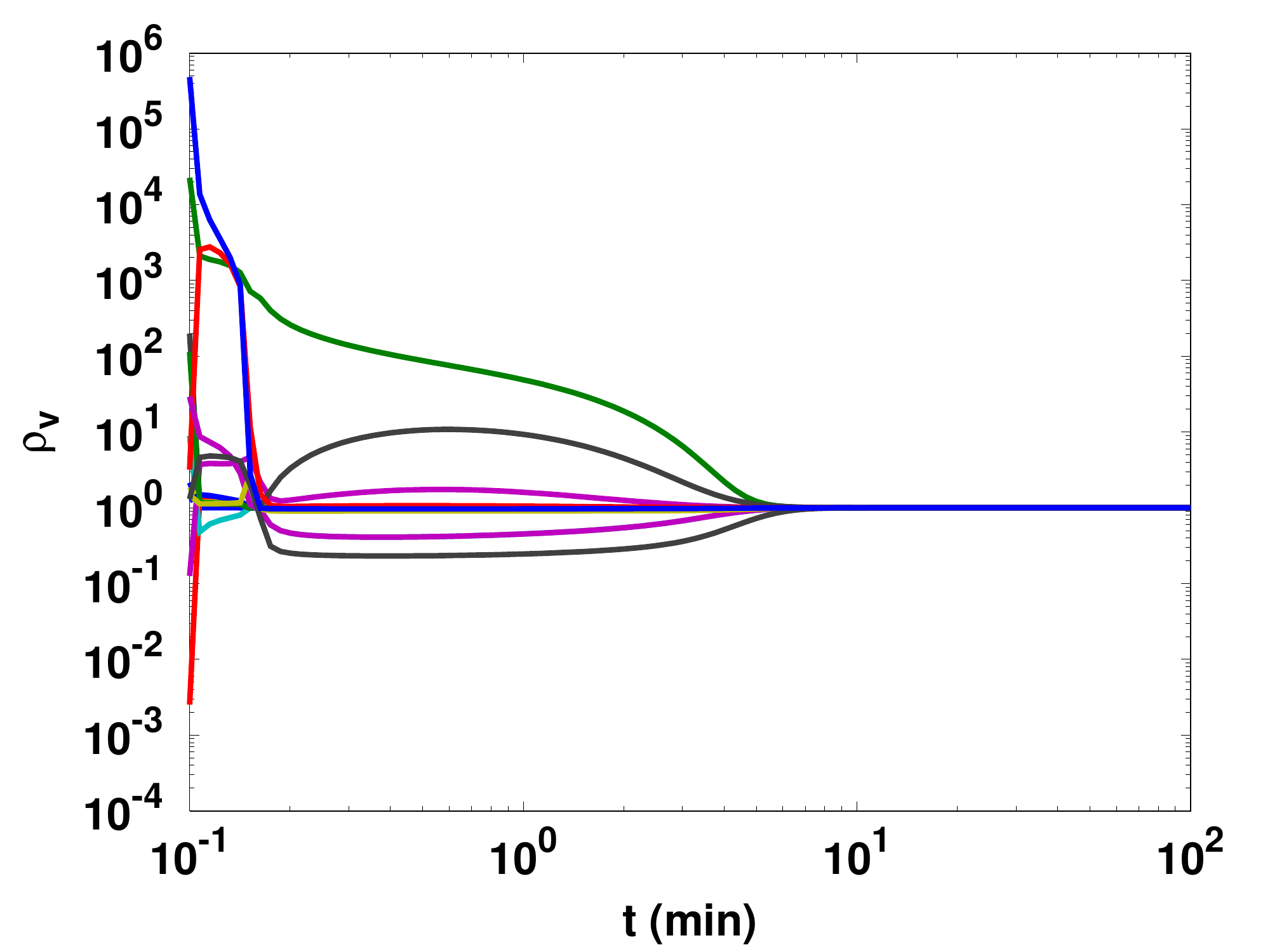}{Closed system}{0.45}
  \SubFig{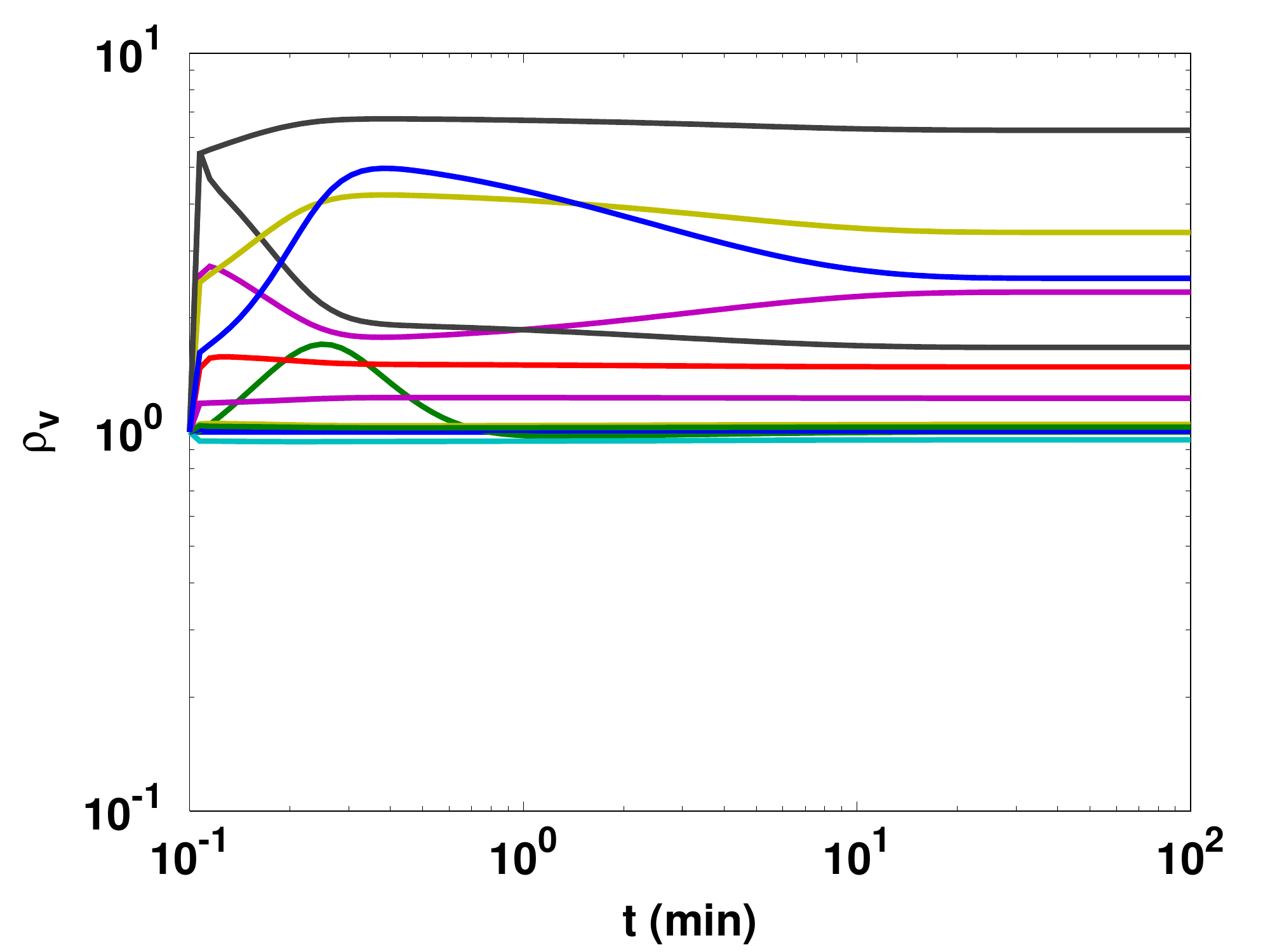}{Open system}{0.45}
  \caption{Simulation: equilibria. For each reaction, the ratio
    $\Vt_0 = K^{v} \hp X^{v}$  \eqref{eq:KvXv} of the forward to backward reaction
    flows, is plotted. (a) Closed system: each ratio tends to unity:
    the steady state is a thermodynamic equilibrium. (b) Open system:
    some ratios tend to a non-unity value: the steady state is a not a
    thermodynamic equilibrium. }
  \label{fig:equilibria}
\end{figure}

\begin{figure}[htbp]
  \centering
  \SubFig{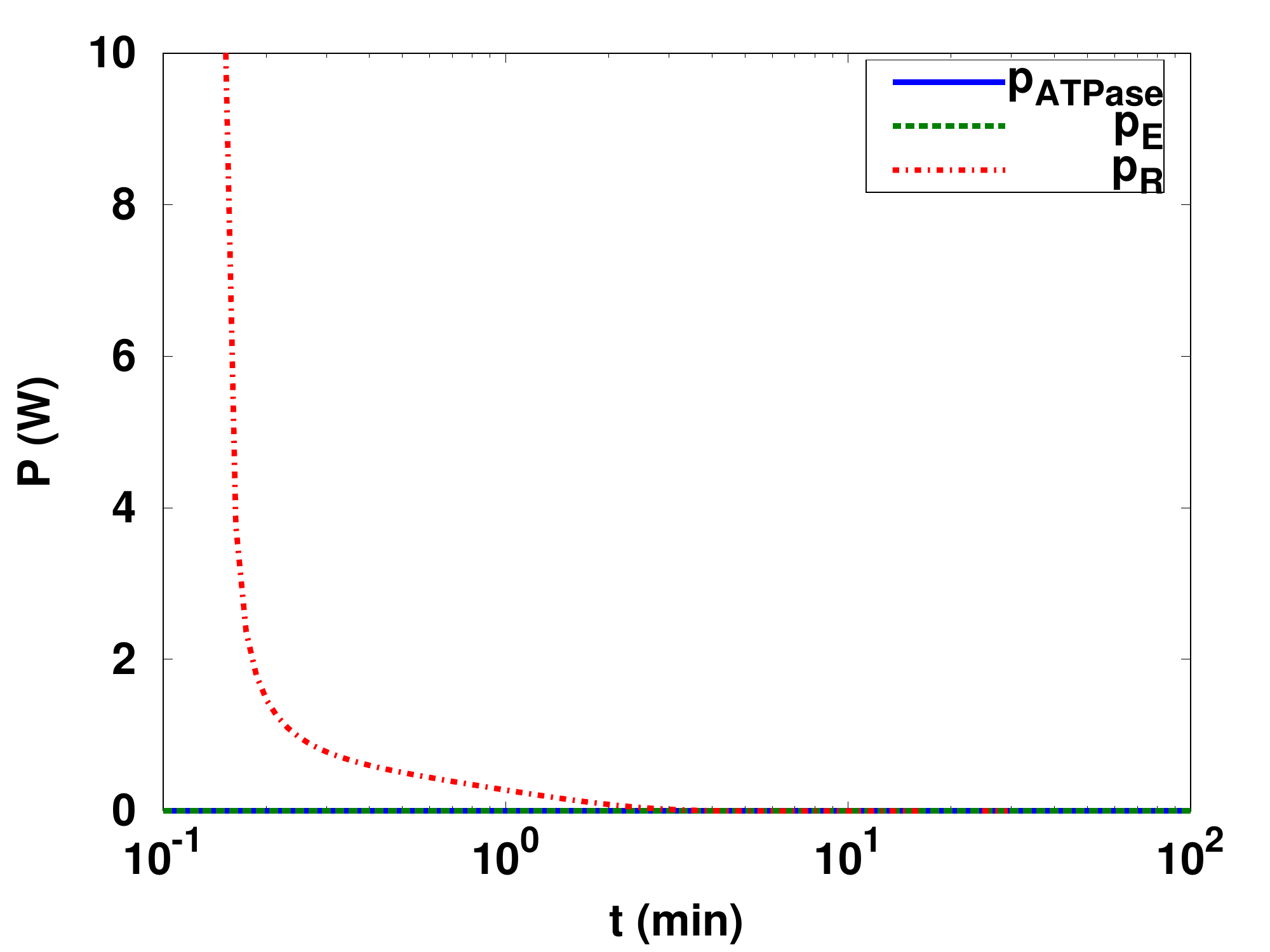}{Closed system}{0.45}
  \SubFig{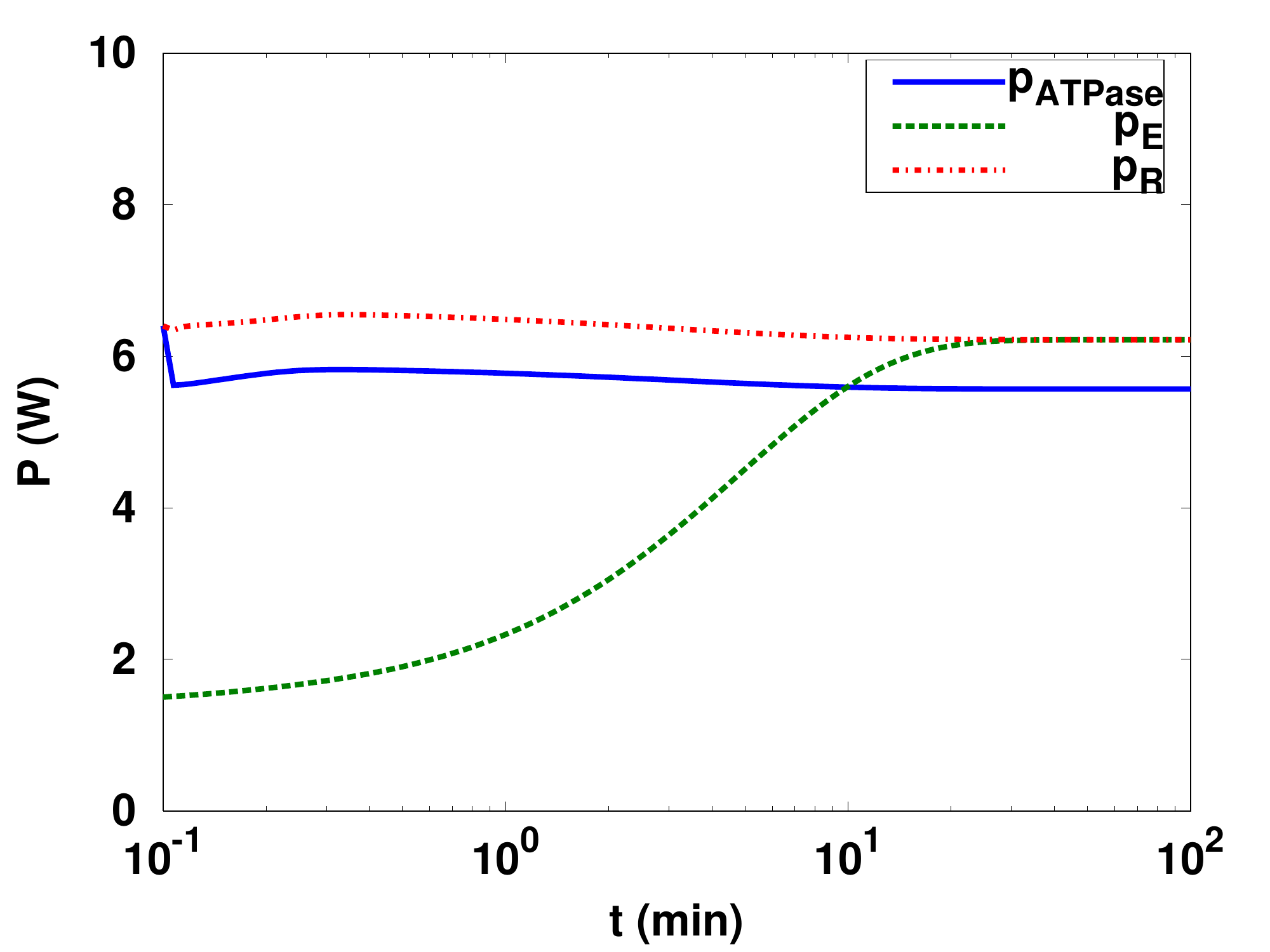}{Open system}{0.45}
  \caption{Simulation: energy flows. Three energy flows are plotted:
    $P_E$ the external energy flow, $P_R$ the energy dissipated in all
    reactions (including $ATPase$), $P_{ATPase}$ the power consumed by
    the processes represented by the $ATPase$ reaction. (a) There is
    no external energy flow or $ATPase$ reaction flow: $P_R$ tends to
    zero as the energy in the internal species is used up. (b) There
    is external energy flow and $ATPase$ reaction flow: $P_R$ tends to
    $P_E$ as the energy in the internal species is used up. In this
    case, at steady-state about 90\% of the energy is associated with
    processes represented by the $ATPase$ reaction.}
  \label{fig:Energy}
\end{figure}

The bond graph  model of Figure \ref{fig:BG} with reaction kinetics
defined by Equation \eqref{eq:v_MM} was compiled into ordinary differential
equations using the bond graph  software MTT (model transformation tools)
\citep{BalBevGawDis05}. Free energy constants were obtained using the
methods of \S~\ref{sec:equil-const} and the
kinetic parameters were derived as in
\S~\ref{sec:enzyme-catal-react}.
The reduced order equations \eqref{eq:reduced}
\S~\ref{app:reduc-order-equat} were simulated using the \texttt{lsode}
solver within GNU Octave \citep{Eat02} numerical software with a
maximum time step of 0.01min
for two cases:
 \paragraph{Closed system.}
  The flows associated with $ATPase$ and $Fout$ were constrained to be
  zero. Initial conditions were defined in \citet[Table 3]{LamKus02},
  together with the extra equations from page 813 describing the redox
  potential (\textasciitilde R) and total creatine abundance, respectively:
  \begin{xalignat}{2}
    \frac{X_{NAD}}{X_{NADH}} &= 1000 &
    X_{Cr} + X_{PCr} &= 40\text{mM}
  \end{xalignat}

  The base units used by \citet{LamKus02} are mM for concentration
  and minutes for time. These units have been used in the following
  figures except that flows are converted to M/s and concentrations to
  M for computing the energy flows in Figure \ref{fig:Energy}.

  
%
  Figure \ref{subfig:LamKus02-0_Xeq} indicates that this steady-state
  is an equilibrium, providing a useful ``thermodynamic validation'' of 
  the model as discussed by \citet{LamKus02}. 
%
  Figure \ref{subfig:LamKus02-0_P} shows energy flows, respectively,
  which become zero after an initial transient as the systems settles
  into equilibrium.

  \paragraph{Open system.}  The flows associated with $ATPase$ and
  $Fout$ were enabled, and the concentration of $LACo$ was fixed at a
  constant small value by adding an appropriate external flow
  $v_E$. The $ATPase$ coefficient was chosen to be 0.75 corresponding
  to the ``Moderate exercise'' column of \citet[Table 4]{LamKus02}.
  The final equilibrium state of the closed system simulation 
  was used as the initial state.
  In contrast to Figure \ref{subfig:LamKus02-0_Xeq}, Figure
  \ref{subfig:LamKus02-4_Xeq} indicates that the steady-state of the
  open system is not an equilibrium as the ratio $\Vt_0$ is not unity.
  Figure \ref{subfig:LamKus02-4_P} shows energy flows which settle to
  non-zero values. In particular, the dissipated power (including
  $ATPase$) $p_R$ becomes equal to the external energy flow
  $P_E$. Energy flows associated with $ATPase$ are indicated
  separately showing that about 90\% of the energy is associated with
  processes represented by the $ATPase$ reaction. The remainder is
  dissipated as heat in the other reactions.

  Further figures appear in the Supplementary material.

\section{Conclusion}
\label{sec:conclusion}





This paper extends the bond graph approach of \citet{GawCra14} to
allow the hierarchical modelling of biochemical systems with reusable
subsystems and robust thermodynamic compliance.
This requires two extensions: the modelling of open thermodynamical
systems using energy ports and the conversion of standard enzymatic
rate parameters to the parameters required by the bond graph.

The reimplimentation of ``A Computational Model for Glycogenolysis in
Skeletal Muscle'', originally presented by \citet{LamKus02}, in a bond
graph formulation verifies the utility of the bond graph approach. In
particular, the discipline imposed by the bond graph reformulation
focuses on four potential problems with the original model: lack of
robustness caused by the separate specification of the identical
equilibrium constant for two parallel reactions ($GPa$ \& $GPb$);
stoichiometric inconsistency arising from equating two versions of the
glycogen molecule; the use of an irreversible reaction ($ATPase$); and
confusion arising from reaction directionality ($TPI$).
A further advantage of bond graph modelling illustrated by this
example is the automatic generation of stoichiometric matrices and
hence conserved moieties. Currently, these must be identified and
imposed by model developers and this becomes a laborious process with
increasing system sizes.

The model of enzymatic reactions used in this paper is that previously
derived by \citet{GawCra14}.
The relationship between this particular bond graph formulation and
some models of enzyme kinetics developed by \citet{LieUhlKli10} is
explored in the paper.  Future work will examine the development of
bond graph representations for a wider range of enzyme models
\citep{Seg93}. This would facilitate the inclusion of allosteric
modulation, such as that exerted by AMP over the $GPb$ reaction.

Metabolic control analysis (MCA) \citep{Fel97} analyses the feedback
control behaviour via sensitivity. There is a well-established
theory of sensitivity bond graphs \citep{Gaw00c,GawRon00b,BorGra02},
which we will use to give a bond graph interpretation of MCA. 
A number of authors have discussed the role of control theory in
systems biology
\citep{TomAce05,WelBulKalMasVer08,IglIng10,CosBat12,CurBal13} and there
is a well-established theory of control in the context of bond graphs
\citep{Kar79a,Gaw95c,GawNeiWag15}. Future work will examine feedback
control of biochemical networks from the bond graph  point of view.
Metabolic networks are further controlled over longer time scales by gene expression 
modulation of maximum reaction rate. Thus, future work will also examine the modulation
of energy flows by gene expression as, for example, found in the
Warburg effect \citep{YizDevRog14}.

An important feature of bond graphs not utilised in this paper is the
ability to interconnect different physical domains. Future work will
examine chemo-electrical and chemo-mechanical transduction as, for
example, found in the cardiac myocyte.

\subsection*{Data accessibility}
A virtual reference environment \citep{HurBudCra14} is available for
this paper at \url{https://sourceforge.net/projects/hbgm/}.
The simulation parameters are listed in the Supplementary material.
\subsection*{Competing interests}
The authors have no competing interests.
\subsection*{Authors' contribution}
ons] All authors contributed to drafting and revising the paper, and
they affirm that they have approved the final version of the
manuscript.
\subsection*{Acknowledgements}
Peter Gawthrop would like to thank the Melbourne School of Engineering
for its support via a Professorial Fellowship.
The authors would like to thank Dr. Ivo Siekmann for alerting them to,
and discussing the contents of, reference \citep{SchRaoJay13} and Dr
Daniel Hurley for creating the virtual reference environment
\citep{HurBudCra14} for this paper.
The authors would like to thank the referees for their encouraging and
insightful comments which lead to an improved manuscript.
\subsection*{Funding statement}
This research was in part conducted and funded by the Australian
Research Council Centre of Excellence in Convergent Bio-Nano Science
and Technology (project number CE140100036), and by the Virtual
Physiological Rat Centre for the Study of Physiology and Genomics,
funded through NIH grant P50-GM094503.


\newpage
\appendix
%

\section{Supplementary Material}
\subsection{Reduced-order equations}
\label{app:reduc-order-equat}
This section includes material from \citet[\S 3(c)]{GawCra14}
about using reduced-order equations for simulation.

Given the reaction flows $V$ of Equation \eqref{eq:open_V}, the rate
of change of the internal states is given by:
\begin{equation}
  \label{eq:dXi}
  \dot{X}_i = N_i V
\end{equation}
As discussed by number of authors the presence of conserved
moieties leads to potential numerical difficulties with the solution of
Equation \eqref{eq:dXi} \citep{Sau09,Ing13}.
Using the notation of \citet[\S 3(c)]{GawCra14}, the reduced-order
state $x$ and the internal state $X_i$ are given by:
\begin{xalignat}{2}
  \label{eq:reduced}
\dot{x} &=  L_{xX}N_iV & X_i = L_{Xx}x + G_X X_i(0)
\end{xalignat}
Equation \eqref{eq:reduced} was used to generate all of the simulation
figures in this paper.

\subsection{Conversion of kinetic data}
\label{app:conv-kinet-data}

\begin{table}[htbp]
  \centering
  \begin{tabular}{|l||l|l|l|l|l|}
    \hline
    Reaction & $K^{eq}$ & $V_{maxf}$ & $V_{maxr}$ & $K^M_a$ & $K^M_b$\\ 
    \hline
    ADK & 2.210e+00 & 8.800e-01 & -- & 8.640e-02& 1.225e-01\\
    CK & 2.330e+02 & 5.000e-01 & -- & 1.330e+01& 1.499e-01\\
    ALD & 9.500e-05 & 1.040e-01 & -- & 5.000e-02& 2.000e+00\\
    TPI & 1.923e+01 & 1.200e+01 & -- & 3.200e-01& 6.100e-01\\
    ENO & 4.900e-01 & 1.920e-01 & -- & 1.000e-01& 3.700e-01\\
    Fout & 1.000e+00 & 2.000e+02 & -- & 1.000e+06& 1.000e+06\\
    PGM & 1.800e-01 & 1.120e+00 & -- & 2.000e-01& 1.400e-02\\
    GAPDH & 8.900e-02 & 1.265e+00 & -- & 6.525e-05& 2.640e-06\\
    LDH & 1.620e+04 & 1.920e+00 & -- & 6.700e-04& 1.443e+01\\
    PGK & 5.711e+04 & 1.120e+00 & -- & 1.600e-05& 4.200e-01\\
    PK & 1.030e+04 & 1.440e+00 & -- & 2.400e-02& 7.966e+00\\
    GPa & 4.200e-01 & 2.000e-02 & -- & 4.000e+00& 2.700e+00\\
    GPb & 4.200e-01 & 3.000e-02 & -- & 2.000e-01& 1.500e+00\\
    PGI & 4.500e-01 & -- & 8.800e-01 & 4.800e-01& 1.190e-01\\
    PGLM & 1.662e+01 & 4.800e-01 & -- & 6.300e-02& 3.000e-02\\
    PFK & 2.420e+02 & 5.600e-02 & -- & 1.440e-02& 1.085e+01\\
    ATPase & 2.497e+05 & 7.500e+02 & -- & 1.000e+06& 1.000e+12\\
    \hline
  \end{tabular}
  \caption{Parameters from \citet[Tables 1\&2]{LamKus02}.}
  \label{tab:LamKus02-par}
\end{table}

\begin{table}[htbp]
  \centering
  \begin{tabular}{|l||l|l|}
\hline
Species & K & $\frac{\mu_0}{\RT}$ \\ 
\hline
ADP & 7.677e-01 & -2.643e-01 \\
AMP & 2.776e-03 & -5.887e+00 \\
ATP & 4.692e+02 & 6.151e+00 \\
Cr & 6.174e-01 & -4.822e-01 \\
P & 2.447e-03 & -6.013e+00 \\
PCr & 1.620e+00 & 4.822e-01 \\
DHAP & 1.038e+00 & 3.718e-02 \\
FBP & 1.968e-03 & -6.231e+00 \\
GAP & 1.996e+01 & 2.994e+00 \\
DPG & 1.512e+06 & 1.423e+01 \\
LAC & 1.748e-18 & -4.089e+01 \\
LACo & 1.748e-18 & -4.089e+01 \\
NAD & 1.659e+03 & 7.414e+00 \\
NADH & 6.026e-04 & -7.414e+00 \\
P2G & 2.406e-01 & -1.425e+00 \\
P3G & 4.330e-02 & -3.140e+00 \\
PEP & 4.910e-01 & -7.114e-01 \\
PYR & 7.795e-08 & -1.637e+01 \\
F6P & 7.792e-04 & -7.157e+00 \\
G1P & 5.827e-03 & -5.145e+00 \\
G6P & 3.506e-04 & -7.956e+00 \\
GLY & 1.000e+00 & 0.000e+00 \\
\hline
  \end{tabular}
  \caption{Bond graph species parameters used in simulation 
    (See \S~\ref{sec:equil-const}}
  \label{tab:species_par}
\end{table}

\begin{table}[htbp]
  \centering
  \begin{tabular}{|l||l|l|l|}
\hline
Reaction & $v_{max}$ & $k_v$ & $\rho_v$\\
\hline
ADK & 3.439e+02 & 4.398e-02 & 6.092e-01\\
CK & 2.418e-02 & 1.863e-01 & 1.000e+00\\
ALD & 1.040e+02 & 9.840e-05 & 2.375e-06\\
TPI & 1.082e+03 & 5.760e-01 & 9.098e-01\\
ENO & 1.695e+02 & 2.124e-02 & 1.169e-01\\
Fout & 1.000e+05 & 8.738e-13 & 5.000e-01\\
PGM & 3.136e+02 & 2.425e-03 & 7.200e-01\\
GAPDH & 3.953e+02 & 1.653e-03 & 6.875e-01\\
LDH & 1.096e+03 & 1.796e-14 & 4.292e-01\\
PGK & 3.527e+02 & 5.847e+00 & 6.851e-01\\
PK & 4.494e+01 & 2.823e-04 & 9.688e-01\\
GPa & 1.233e+01 & 6.035e-03 & 3.836e-01\\
GPb & 2.841e+01 & 4.635e-04 & 5.303e-02\\
PGI & 5.674e+02 & 5.978e-05 & 6.448e-01\\
PGLM & 1.337e+01 & 1.023e-05 & 9.721e-01\\
PFK & 4.239e+01 & 3.985e-03 & 2.430e-01\\
ATPase & 6.001e+05 & 3.755e+08 & 1.998e-01\\
\hline
  \end{tabular}
  \caption{Bond graph reaction parameters used in simulation 
    (See \S~\ref{sec:enzyme-catal-react}}
  \label{tab:reaction_par}
\end{table}

Using the methods of
\S~\ref{sec:equil-const}, the equilibrium
constants quoted by \citet[Tables 1\&2]{LamKus02} (Table \ref{tab:LamKus02-par}) were converted into the
free-energy constants required by the bond graph formulation and are
listed in Table \ref{tab:species_par}.

Using the methods of
\S~\ref{sec:enzyme-catal-react}, the
reaction constants 
quoted by \citet{LamKus02} (Table \ref{tab:LamKus02-par}) were
converted into the reaction constants required by the bond graph
formulation of \S~\ref{sec:enzyme-catal-react} and are listed in
Table \ref{tab:reaction_par}.

Section \ref{app:mass-acti-react} looks at mass-action reactions as
used for $ATPase$, \S~\ref{sec:relat-direct-bind} looks at the
relationship of the approach in
\S~\ref{sec:enzyme-catal-react} to the
direct-binding  modular rate law of
  \citet{LieUhlKli10}, \S~\ref{sec:relat-comm-modul} to the common modular rate law of
  \citet{LieUhlKli10} and \S~\ref{sec:relat-comp-model} to the
  computational model of \citet{LamKus02}.

\subsection{Mass-action reactions}
\label{app:mass-acti-react}
The mass-action formulation of chemical equations reveals key issues
encountered in converting kinetic data from enzymatic models into the
form required by a bond graph model. Enzyme catalysed reactions are
discussed in \S~\ref{sec:enzyme-catal-react}.

The mass-action formulation presented by \citet[Equation 2.6]{GawCra14} uses the
\emph{Marcelin} formulation rewritten here as:
\begin{xalignat}{3}
  v &= \kappa \lb V_0^+ - V_0^- \rb&
  \text{where }
  V_0^+ &= e^{A^f/RT}&
  \text{and }
  V_0^- &= e^{A^r/RT}
\end{xalignat}
In terms of the reaction $A \reac B$
\begin{xalignat}{2}
  V_0^+ &= K_a x_a & V_0^- &= K_b x_b
\end{xalignat}
In terms of the reaction $A+B \reac 2C$
\begin{xalignat}{2}
  V_0^+ &= K_a x_a K_b x_b & V_0^- &= \lb K_c x_c \rb^2
\end{xalignat}

One standard way of writing the Mass-action rate of $A \reac B$
\begin{xalignat}{2}
  v &= \kappa_{eq} \lb x_a - \frac{1}{K_{eq}} x_b \rb&
\text{where }
K_{eq} &= \frac{K_a}{K_b}
\end{xalignat}
similarly, $A+B \reac 2C$ can be rewritten as
\begin{xalignat}{2}
  v &= \kappa_{eq} \lb x_ax_b - \frac{1}{K_{eq}} x_c^2 \rb&
\text{where }
K_{eq} &= \frac{K_aK_b}{K_c^2}
\end{xalignat}
Define $K^f$ as the constant on the substrate side and $K^r$ as the
constant on the product side. In the case of $A \reac B$
\begin{xalignat}{2}
  K^f &= K_a & \text{and } K^r &= K_b
\end{xalignat}
and in the case of $A+B \reac 2C$
\begin{xalignat}{2}
  K^f &= K_aK_b & \text{and } K^r &= K_c^2
\end{xalignat}
It follows that:
\begin{xalignat}{2}
  \kappa_{eq} &=  K^f \kappa & \text{and } K_{eq} &= \frac{K^f}{K_r}
\end{xalignat}

As $K_f$ can be computed from $K$, which in turn can be deduced as
discussed in \S~\ref{sec:equil-const}, it follows 
that $\kappa$ can be deduced from $\kappa_{eq}$.
\subsection{Relation to the Direct Binding Modular Rate Law of
  \citet{LieUhlKli10}}
\label{sec:relat-direct-bind}
It is now shown using an example that Equation \eqref{eq:v_MM} is of
the same form as the \emph{direct binding modular} rate law of
\citet{LieUhlKli10}.  Consider the enzyme catalysed reaction $A+B
\reac 2C$ In this case:
\begin{equation}
  V_0^+ = K_a K_b x_a x_b,\; V_0^- = K_c^2  x_c^2
\end{equation}
Equation \eqref{eq:v_MM} is of the form:
\begin{equation}
  \label{eq:v_mm_AB2C}
    v = \frac
  {v_{max}^+\frac{K_a K_b x_a x_b}{K_v^+} - v_{max}^-\frac{K_c^2 x_c^2}{K_v^-}}
  {1 + \frac{K_a K_b x_a x_b}{K_v^+} + \frac{K_c^2 x_c^2}{K_v^-}}
\end{equation}
The \emph{direct binding modular} rate law is given by Equation (4) of \citet{LieUhlKli10} 
and in the notation of this paper is:
\begin{equation}
  \label{eq:DM}
  v = u \frac{k^+\frac{x_a}{\bK_a}\frac{x_b}{\bK_b} -
    k^-\lb\frac{x_c}{\bK_c}\rb^2}
{1 + \frac{x_a}{\bK_a}\frac{x_b}{\bK_b} + \lb\frac{x_c}{\bK_c}\rb^2}
\end{equation}
Equations \eqref{eq:v_mm_AB2C} and \eqref{eq:DM} are identical if we set:
\begin{xalignat}{4}
  v_{max}^+ &= uk^+ & 
  v_{max}^- &= uk^- &  
  K_v^+ &= \frac{\bK_a\bK_b}{K_a K_b} &
  K_v^- &= \lb\frac{\bK_c}{K_c}\rb^2
\end{xalignat}

\subsection{Relation to the Common Modular Rate Law of
  \citet{LieUhlKli10}}\label{sec:relat-comm-modul}
However, Equation \eqref{eq:v_MM} is not the same as the \emph{common modular}
rate law of \citet{LieUhlKli10}.
In the context of the reaction $A+B \reac 2C$, the \emph{common modular}
rate law of \citet{LieUhlKli10} is of the form:
\begin{equation}
  \label{eq:CM}
  v = u \frac{k^+\frac{x_a}{\bK_a}\frac{x_b}{\bK_b} -
    k^-\lb\frac{x_c}{\bK_c}\rb^2}
{1 + \frac{x_a}{\bK_a} + \frac{x_b}{\bK_b} +
  \frac{x_a}{\bK_a}\frac{x_b}{\bK_b} + 2\frac{x_c}{\bK_c}
+ \lb\frac{x_c}{\bK_c}\rb^2}
\end{equation}
As discussed by \citet{LieUhlKli10}, the additional
denominator terms imply that Equation \eqref{eq:CM} is not the same as Equation
\eqref{eq:DM} but can be considered an approximation to it. However,
in the case of reaction $A \reac B$, the common modular and direct binding modular reaction rates
are identical.

\subsection{Relation to the computational model of \citet{LamKus02}}
\label{sec:relat-comp-model}
The general enzyme catalysed reaction between two species is given by
$A \reac B$ and the corresponding rate
is written by \citet{LamKus02} as:
\begin{equation}
  \label{eq:v_LamKus02_AB}
  v =  \frac{V_{maxf} \frac{x_a}{\bK_{a}} 
    - V_{maxr}\frac{x_b}{\bK_{b}} }
  {1 + \frac{x_a}{\bK_{a}} + \frac{x_b}{\bK_{b}}}
\text{ where } V_{maxr} = V_{maxf}\frac{\bK_{b}\bK_{eq}}{\bK_{a}}
\end{equation}
In this case $V_0^+=K_a X_a$ and $V_0^-=K_b X_b$ hence Equation
\eqref{eq:v_MM} becomes:
 \begin{equation}
  \label{eq:v_MM_AB}
  v = \frac{v_{max}}{k_v} \frac{K_a X_a - K_b X_b}{1 + 
    \lb \lb 1 - \rho_v \rb \frac{K_a}{k_v} X_a + \rho_v  \frac{K_b}{k_v} X_b\rb}
\end{equation}
Comparing Equations \eqref{eq:v_LamKus02_AB} and \eqref{eq:v_MM_AB}
and using Equation \eqref{eq:v_max_rho}
it follows that they are identical if:
\begin{equation}
  \bK_{a} = \frac{1}{1-\rho_v}\frac{k_v}{K_a} \text { and } \bK_{b} = \frac{1}{\rho_v}\frac{k_v}{K_b}
\end{equation}
or:
\begin{equation}
  k_v = (1-\rho_v) K_a \bK_a \text { and } k_v = \rho_v K_b \bK_b
\end{equation}
Having deduced $\rho_v$ from the given data using Equation
\eqref{eq:v_max_rho}, $k_v$ can then be deduced from $\bK_a$ using
$K_a$, as shown in \S~\ref{sec:equil-const}.

Alternatively comparing Equation \eqref{eq:v_MM_0} with Equation
\eqref{eq:v_LamKus02_AB} gives:
\begin{equation}
  k_v^+ = \frac{\bK_{a}}{K_a} \text { and }
  k_v^- = \frac{\bK_{b}}{K_b}
\end{equation}
Using the expressions for $k_v^+$ and $k_v^-$ \eqref{eq:v_MM_0} gives
the same result.

\subsection{TPI}
\label{app:TPI}
The equilibrium constant is given as $K_{eq} =  0.052$. However this
gives the wrong value of $K_{eqcombo}$. Because the reaction is
specified in the ``wrong'' direction, it is assumed that $K_{eq}$
should be the reciprocal of the given value, ie $K_{eq} = 19.23$.
\citet{Bea12} quotes $K_{eq} = 19.87$; so this alteration seems to be
correct.

\subsection{Hierarchical modelling}
\label{app:hier-modell}
The bond graphs of the subsystems \textbf{GAP2LAC} and \textbf{CrAMP}
appear in Figures \ref{fig:GAP2LAC} and \ref{fig:CrAMP}.
%
\SubModel{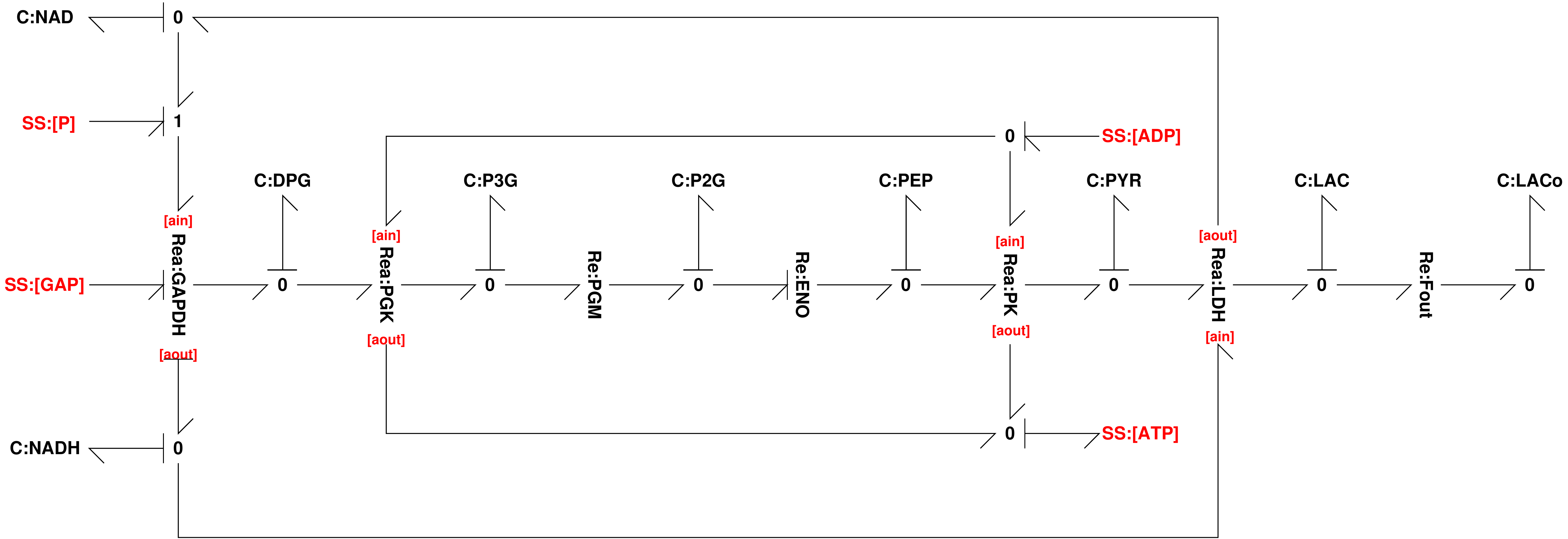}{0.9}
\SubModel{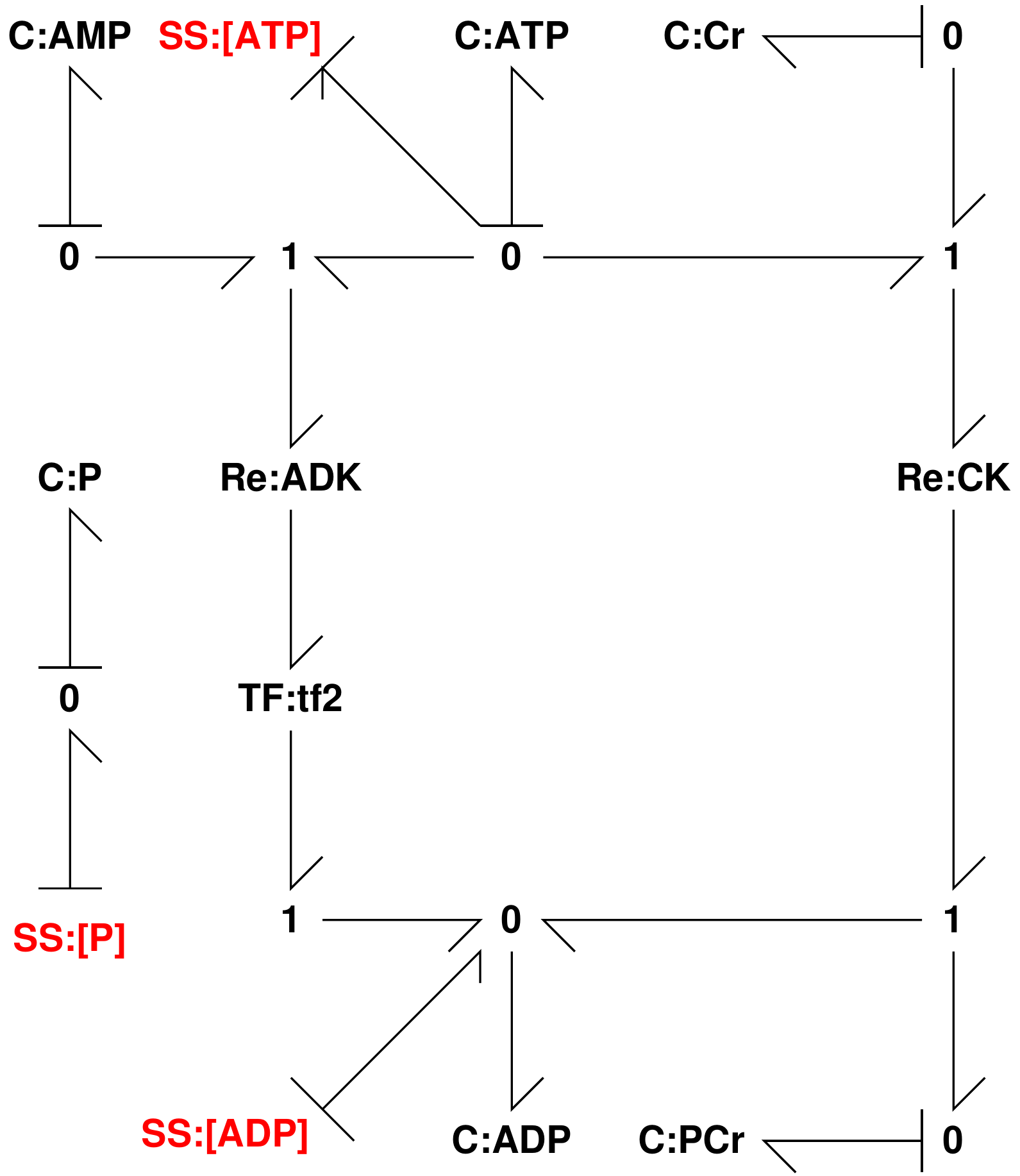}{0.4}
%

The ODEs, and corresponding flows, automatically generated from the
Bond Graph are given by the following equations

\begin{equation}
\begin{aligned}
\dot X_{adp} &=
{
2 V_{adk} - V_{pgk} - V_{pk} + V_{pfk} + V_{atpase} + V_{ck}
}
\cr
\dot X_{amp} &=
{
 - V_{adk}
}
\cr
\dot X_{atp} &=
{
 - V_{adk} + V_{pgk} + V_{pk} - V_{pfk} - V_{atpase} - V_{ck}
}
\cr
\dot X_{cr} &=
{
 - V_{ck}
}
\cr
\dot X_{p} &=
{
 - V_{gpa} - V_{gpb} + V_{atpase} - V_{gapdh}
}
\cr
\dot X_{pcr} &=
{
V_{ck}
}
\cr
\dot X_{dhap} &=
{
V_{ald} + V_{tpi}
}
\cr
\dot X_{fbp} &=
{
V_{pfk} - V_{ald}
}
\cr
\dot X_{gap} &=
{
V_{ald} - V_{tpi} - V_{gapdh}
}
\cr
\dot X_{dpg} &=
{
 - V_{pgk} + V_{gapdh}
}
\cr
\dot X_{lac} &=
{
 - V_{fout} + V_{ldh}
}
\cr
\dot X_{laco} &=
{
V_{fout}
}
\cr
\dot X_{nad} &=
{
 - V_{gapdh} + V_{ldh}
}
\cr
\dot X_{nadh} &=
{
V_{gapdh} - V_{ldh}
}
\cr
\dot X_{p2g} &=
{
 - V_{eno} + V_{pgm}
}
\cr
\dot X_{p3g} &=
{
V_{pgk} - V_{pgm}
}
\cr
\dot X_{pep} &=
{
 - V_{pk} + V_{eno}
}
\cr
\dot X_{pyr} &=
{
V_{pk} - V_{ldh}
}
\cr
\dot X_{f6p} &=
{
V_{pgi} - V_{pfk}
}
\cr
\dot X_{g1p} &=
{
V_{gpa} + V_{gpb} - V_{pglm}
}
\cr
\dot X_{g6p} &=
{
 - V_{pgi} + V_{pglm}
}
\cr
\dot X_{gly} &=
{
0
}
\end{aligned}
\end{equation}
\begin{equation}
\begin{aligned}
V_{adk} &=
{
{\left (v_{adk} {\left ( - k_{adp}^2 X_{adp}^2 + k_{amp} k_{atp} X_{amp} X_{atp}\right )}\right )} \over {\left (k_{adk} + k_{adp}^2 X_{adp}
^2 \rho_{adk} - k_{amp} k_{atp} X_{amp} X_{atp} \rho_{adk} + k_{amp} k_{atp} X_{amp} X_{atp}\right )}
}
\cr
V_{ck} &=
{
{\left (v_{ck} {\left ( - k_{adp} k_{pcr} X_{adp} X_{pcr} + k_{atp} k_{cr} X_{atp} X_{cr}\right )}\right )} \over {\left (k_{adp} k_{pcr} X_{adp}
X_{pcr} \rho_{ck} - k_{atp} k_{cr} X_{atp} X_{cr} \rho_{ck} + k_{atp} k_{cr} X_{atp} X_{cr} + k_{ck}\right )}
}
\cr
V_{ald} &=
{
{\left (v_{ald} {\left ( - k_{dhap} k_{gap} X_{dhap} X_{gap} + k_{fbp} X_{fbp}\right )}\right )} \over {\left (k_{ald} + k_{dhap} k_{gap} X_{dhap}
X_{gap} \rho_{ald} - k_{fbp} X_{fbp} \rho_{ald} + k_{fbp} X_{fbp}\right )}
}
\cr
V_{tpi} &=
{
{\left (v_{t\pi} {\left ( - k_{dhap} X_{dhap} + k_{gap} X_{gap}\right )}\right )} \over {\left (k_{dhap} X_{dhap} \rho_{t\pi} - k_{gap} X_{gap}
\rho_{t\pi} + k_{gap} X_{gap} + k_{t\pi}\right )}
}
\cr
V_{eno} &=
{
{\left (v_{eno} {\left (k_{p2g} X_{p2g} - k_{pep} X_{pep}\right )}\right )} \over {\left (k_{eno} - k_{p2g} X_{p2g} \rho_{eno} + k_{p2g}
X_{p2g} + k_{pep} X_{pep} \rho_{eno}\right )}
}
\cr
V_{fout} &=
{
{\left (v_{fout} {\left (k_{lac} X_{lac} - k_{laco} X_{laco}\right )}\right )} \over {\left (k_{fout} - k_{lac} X_{lac} \rho_{fout} + k_{lac}
X_{lac} + k_{laco} X_{laco} \rho_{fout}\right )}
}
\cr
V_{pgm} &=
{
{\left (v_{pgm} {\left ( - k_{p2g} X_{p2g} + k_{p3g} X_{p3g}\right )}\right )} \over {\left (k_{p2g} X_{p2g} \rho_{pgm} - k_{p3g} X_{p3g}
\rho_{pgm} + k_{p3g} X_{p3g} + k_{pgm}\right )}
}
\cr
V_{gapdh} &=
{
{\left (v_{gapdh} {\left ( - k_{dpg} k_{nadh} X_{dpg} X_{nadh} + k_{gap} k_{nad} k_{p} X_{nad} X_{p} X_{gap}\right )}\right )} \over {\left (
k_{dpg} k_{nadh} X_{dpg} X_{nadh} \rho_{gapdh} - k_{gap} k_{nad} k_{p} X_{nad} X_{p} X_{gap}
\rho_{gapdh} + k_{gap} k_{nad} k_{p} X_{nad} X_{p} X_{gap} + k_{gapdh}\right )}
}
\cr
V_{ldh} &=
{
{\left (v_{ldh} {\left ( - k_{lac} k_{nad} X_{lac} X_{nad} + k_{nadh} k_{pyr} X_{nadh} X_{pyr}\right )}\right )} \over {\left (k_{lac} k_{nad}
 X_{lac} X_{nad} \rho_{ldh} + k_{ldh} - k_{nadh} k_{pyr} X_{nadh} X_{pyr} \rho_{ldh} + k_{nadh}
k_{pyr} X_{nadh} X_{pyr}\right )}
}
\cr
V_{pgk} &=
{
{\left (v_{pgk} {\left ( - k_{adp} k_{dpg} X_{adp} X_{dpg} + k_{atp} k_{p3g} X_{p3g} X_{atp}\right )}\right )} \over {\left (k_{adp} k_{dpg}
X_{adp} X_{dpg} \rho_{pgk} - k_{adp} k_{dpg} X_{adp} X_{dpg} - k_{atp} k_{p3g} X_{p3g} X_{atp}
\rho_{pgk} - k_{pgk}\right )}
}
\cr
V_{pk} &=
{
{\left (v_{pk} {\left ( - k_{adp} k_{pep} X_{adp} X_{pep} + k_{atp} k_{pyr} X_{pyr} X_{atp}\right )}\right )} \over {\left (k_{adp} k_{pep}
X_{adp} X_{pep} \rho_{pk} - k_{adp} k_{pep} X_{adp} X_{pep} - k_{atp} k_{pyr} X_{pyr} X_{atp} \rho_{pk}
 - k_{pk}\right )}
}
\cr
V_{gpa} &=
{
{\left (k_{gly} X_{gly} v_{gpa} {\left ( - k_{g1p} X_{g1p} + k_{p} X_{p}\right )}\right )} \over {\left (k_{g1p} k_{gly} X_{g1p} X_{gly}
\rho_{gpa} - k_{gly} k_{p} X_{gly} X_{p} \rho_{gpa} + k_{gly} k_{p} X_{gly} X_{p} + k_{gpa}\right )}
}
\cr
V_{gpb} &=
{
{\left (k_{gly} X_{gly} v_{gpb} {\left ( - k_{g1p} X_{g1p} + k_{p} X_{p}\right )}\right )} \over {\left (k_{g1p} k_{gly} X_{g1p} X_{gly}
\rho_{gpb} - k_{gly} k_{p} X_{gly} X_{p} \rho_{gpb} + k_{gly} k_{p} X_{gly} X_{p} + k_{gpb}\right )}
}
\cr
V_{pgi} &=
{
{\left (v_{pgi} {\left ( - k_{f6p} X_{f6p} + k_{g6p} X_{g6p}\right )}\right )} \over {\left (k_{f6p} X_{f6p} \rho_{pgi} - k_{g6p} X_{g6p}
\rho_{pgi} + k_{g6p} X_{g6p} + k_{pgi}\right )}
}
\cr
V_{pglm} &=
{
{\left (v_{pglm} {\left ( - k_{g1p} X_{g1p} + k_{g6p} X_{g6p}\right )}\right )} \over {\left (k_{g1p} X_{g1p} \rho_{pglm} - k_{g1p} X_{g1p}
- k_{g6p} X_{g6p} \rho_{pglm} - k_{pglm}\right )}
}
\cr
V_{pfk} &=
{
{\left (v_{pfk} {\left ( - k_{adp} k_{fbp} X_{adp} X_{fbp} + k_{atp} k_{f6p} X_{f6p} X_{atp}\right )}\right )} \over {\left (k_{adp} k_{fbp}
X_{adp} X_{fbp} \rho_{pfk} - k_{atp} k_{f6p} X_{f6p} X_{atp} \rho_{pfk} + k_{atp} k_{f6p} X_{f6p}
X_{atp} + k_{pfk}\right )}
}
\cr
V_{atpase} &=
{
{\left (v_{atpase} {\left ( - k_{adp} k_{p} X_{adp} X_{p} + k_{atp} X_{atp}\right )}\right )} \over {\left (k_{adp} k_{p} X_{adp} X_{p}
\rho_{atpase} - k_{atp} X_{atp} \rho_{atpase} + k_{atp} X_{atp} + k_{atpase}\right )}
}
\end{aligned}
\end{equation}

\subsection{Simulation}
\label{app:sim}
\begin{figure}[htbp]
  \centering
  \SubFig{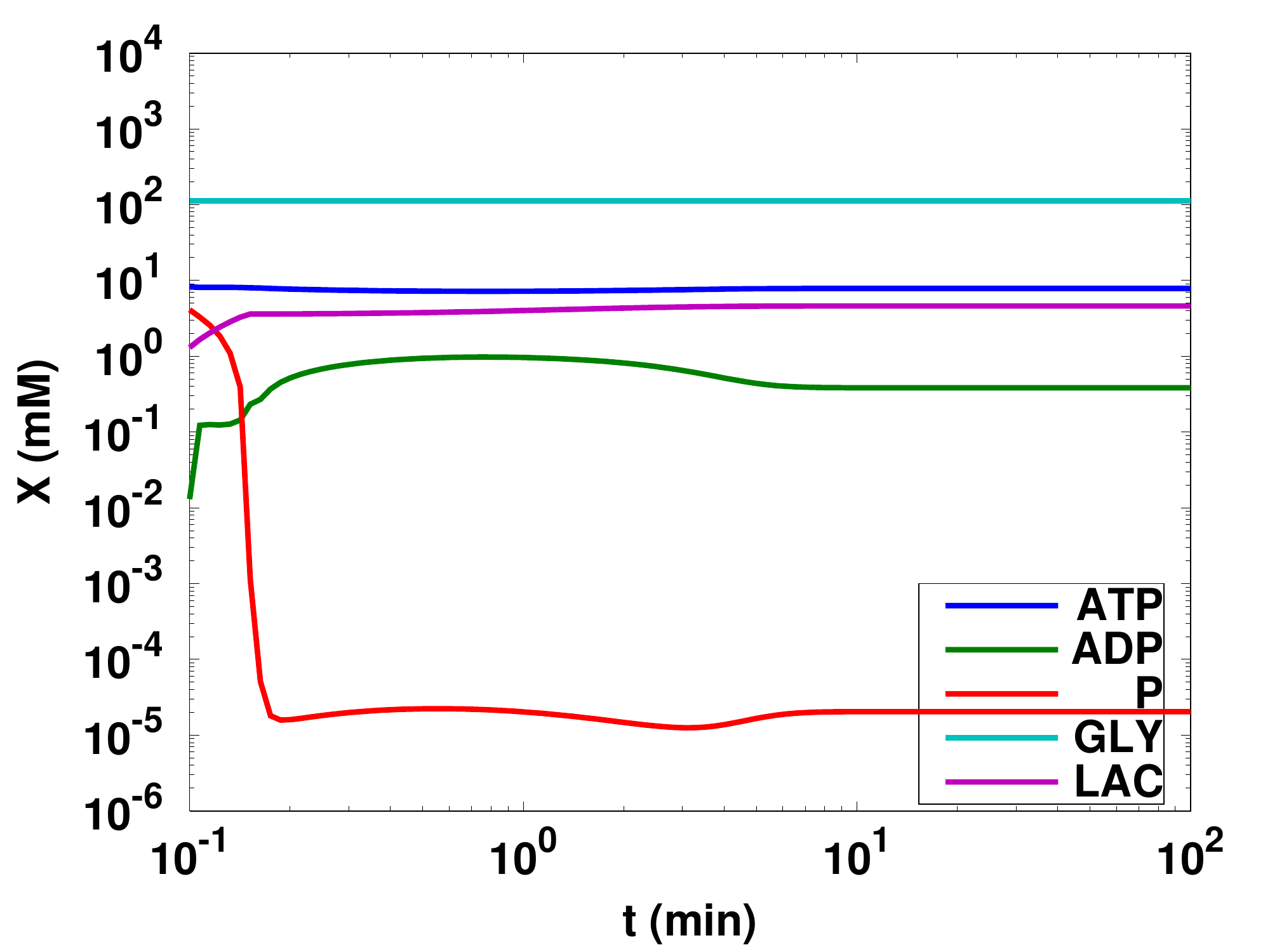}{Closed system}{0.45}
  \SubFig{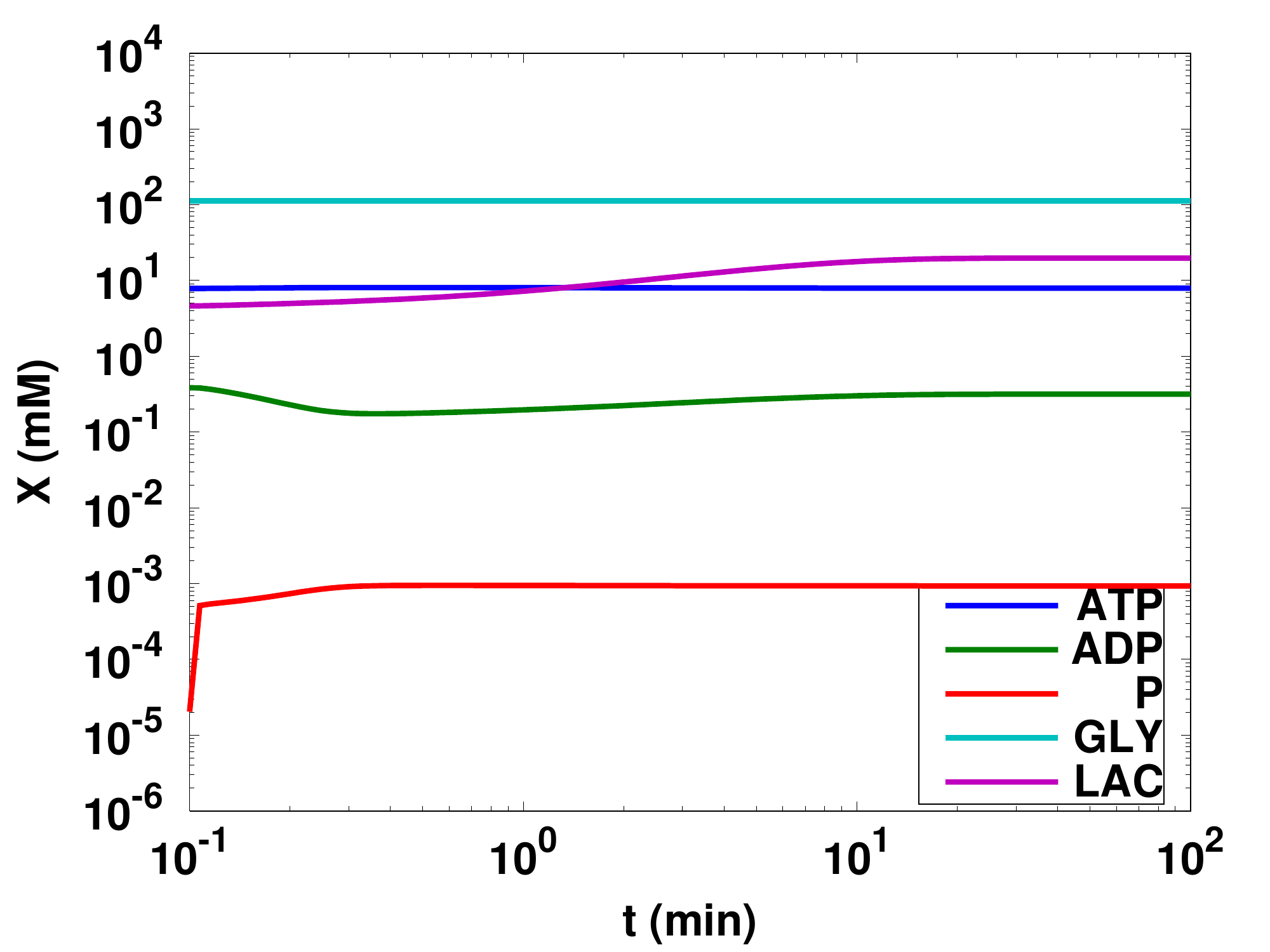}{Open system}{0.45}
  \caption{Simulated concentrations.  Evolution of the concentrations
    for $ATP$, $ADP$, $P$, $GLY$ and $LAC$ corresponds to the
    situation in \citet[Figure 2]{LamKus02} except that they use unit
    initial states. Following an initial transient, the species
    concentrations reach steady-state values for both the closed and
    open systems (cf Figure \ref{fig:sim_ABCAo}).}
  \label{fig:concentrations}
\end{figure}

\begin{figure}[htbp]
  \centering
  \SubFig{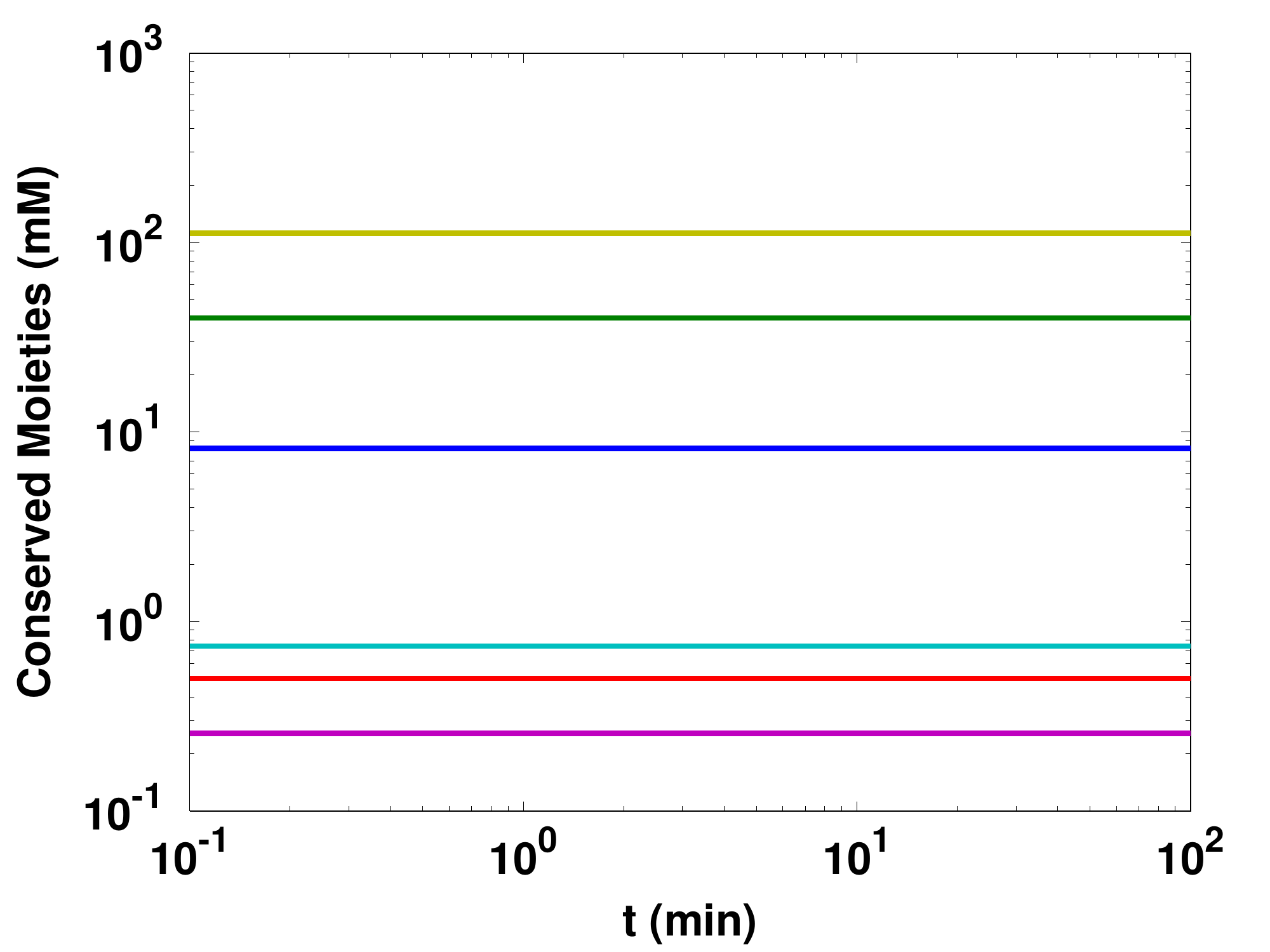}{Closed system}{0.45}
  \SubFig{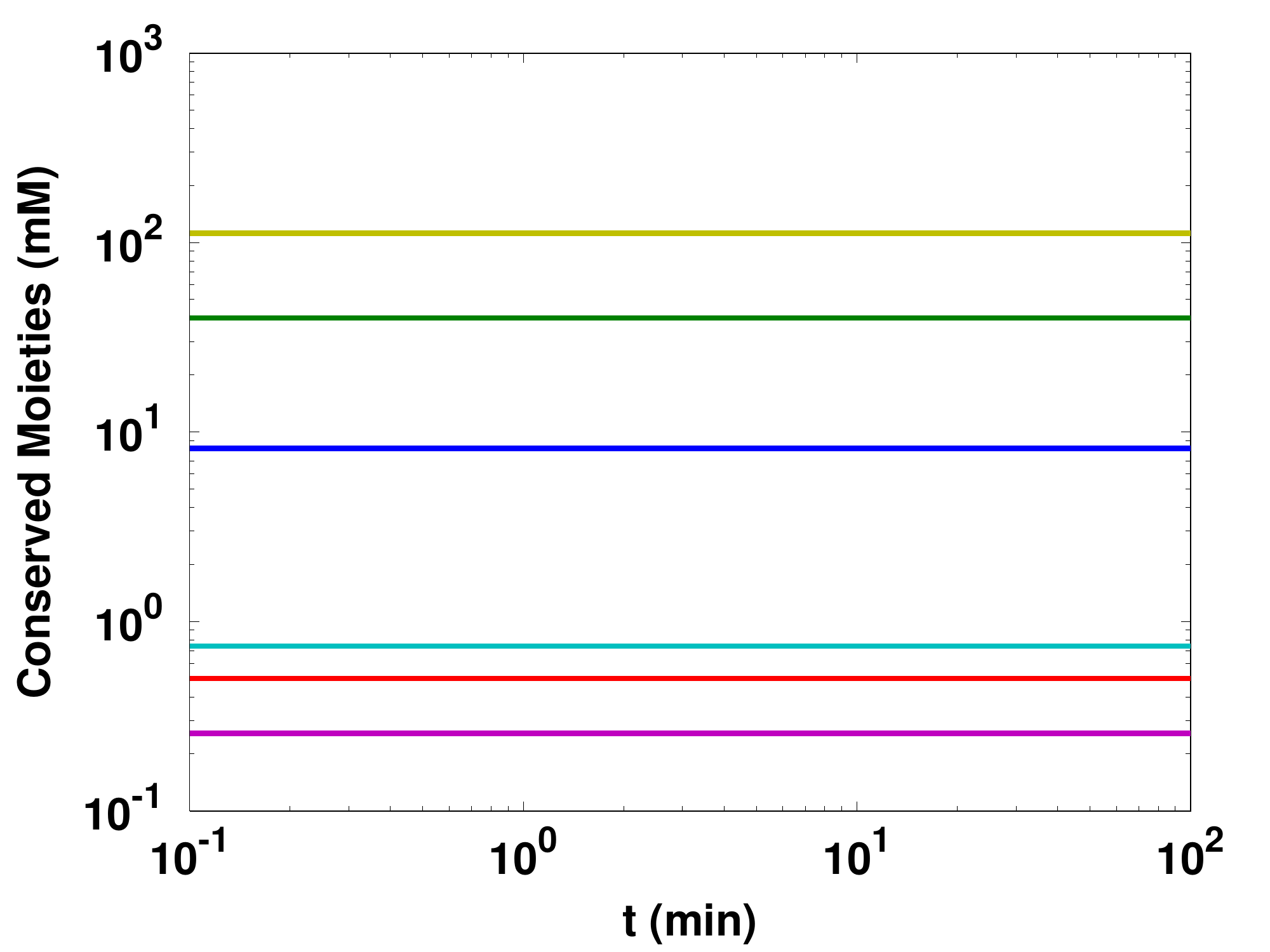}{Open system}{0.45}
  \caption{Conserved Moieties. The sum of each conserved moiety
    remains constant.}
  \label{fig:moieties}
\end{figure}

\begin{figure}[htbp]
  \centering
  \SubFig{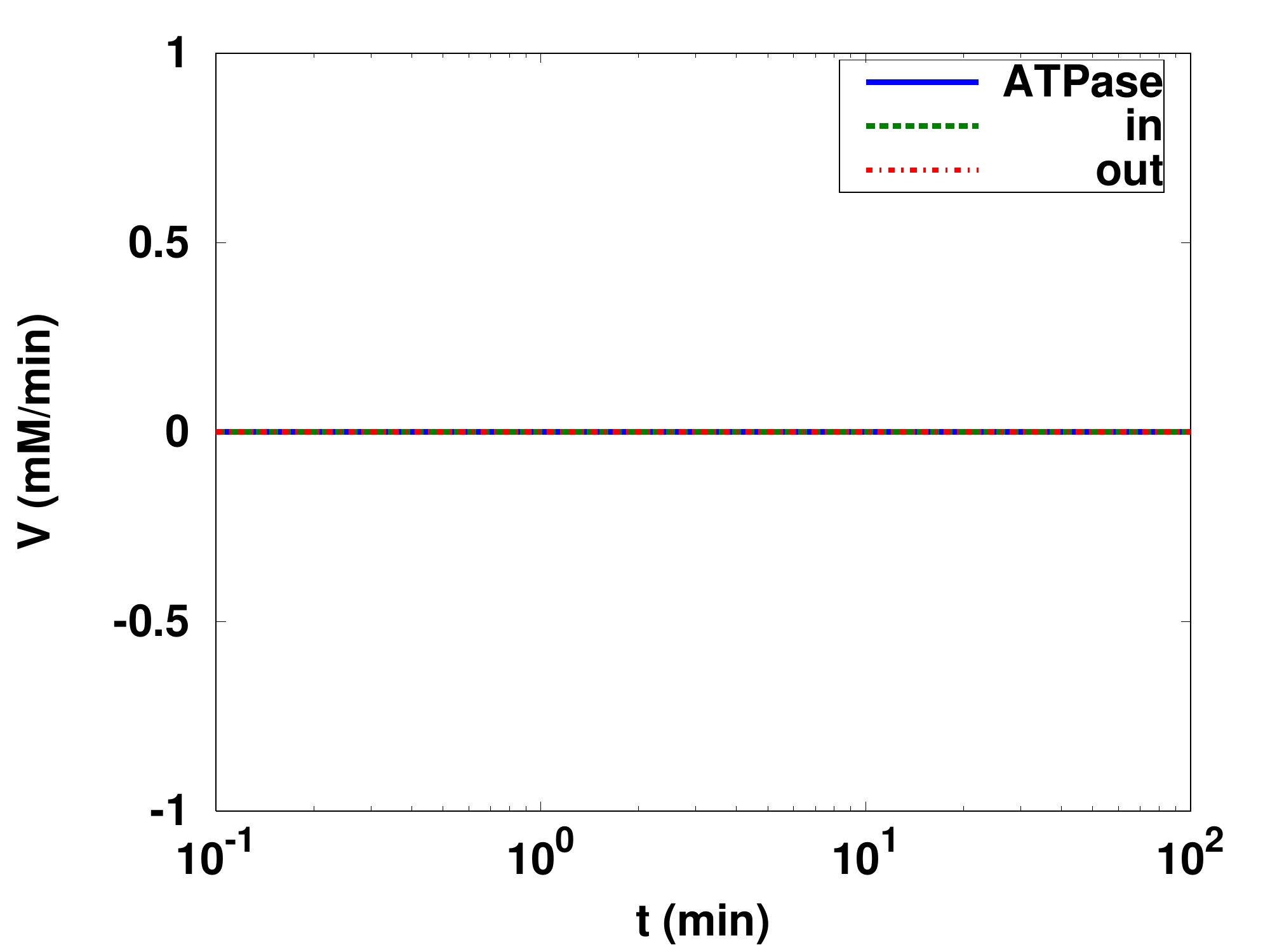}{Closed system}{0.45}
  \SubFig{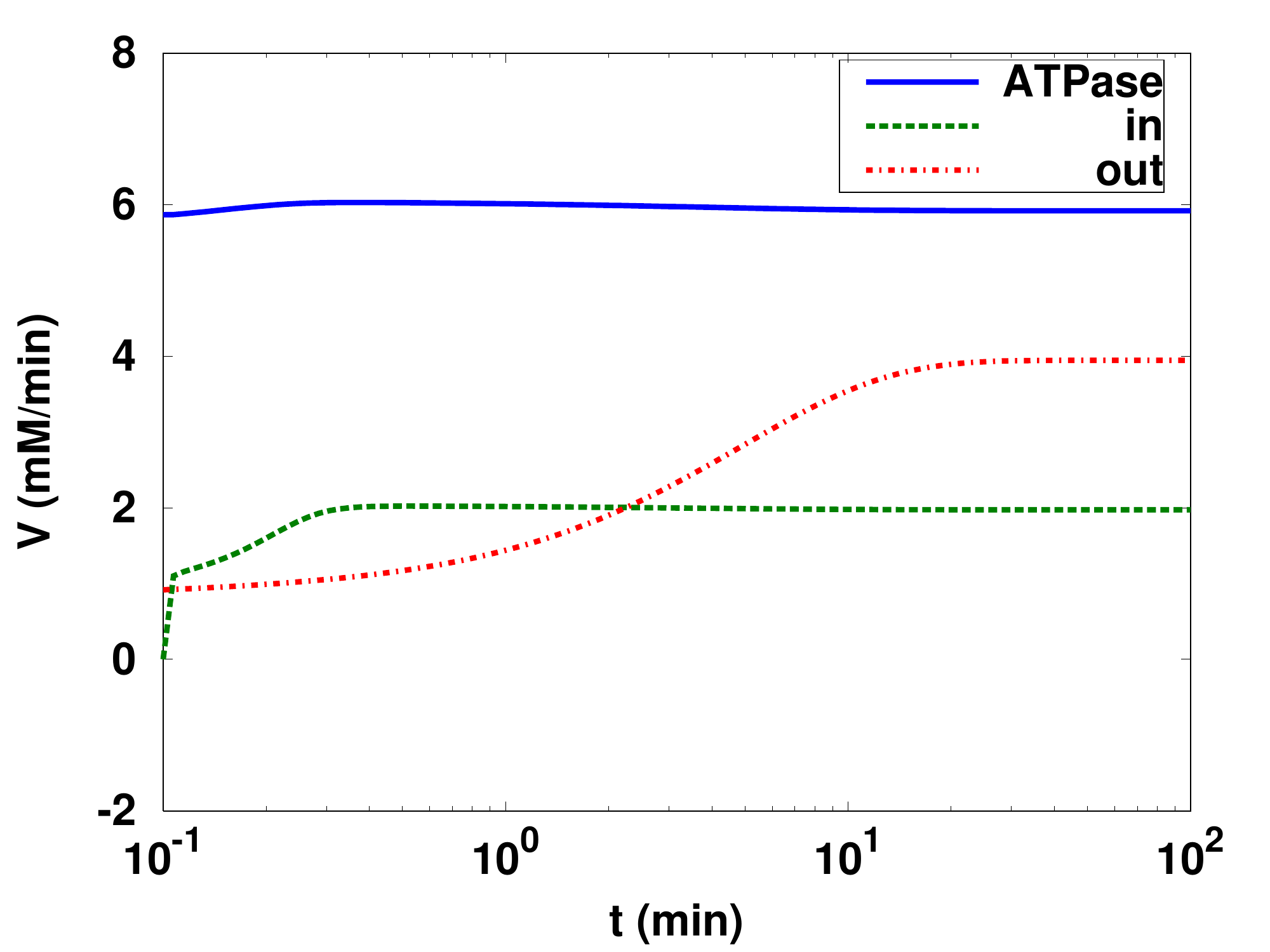}{Open system}{0.45}
  \caption{Simulated mass flows. Both plots show the input mass flow
    (though \textbf{SS:GLYo}), the output mass flow (though
    \textbf{Re:Fout}) and the ATP flux (though \textbf{Re:ATPase}) (a)
    In the closed system, the flows are zero. (b) In the open system,
    the three flows reach a constant steady state. The final value of
    ATP flow is 5.9mM/min; this is close to to value of
    6.1mM/min quoted in the ``Moderate exercise'' column
    of \citet[Table 4]{LamKus02}}
  \label{fig:mass}
\end{figure}

Further figures corresponding to Section \ref{sec:sim} appear in
Figures  \ref{fig:concentrations},
\ref{fig:moieties} and \ref{fig:mass}.

\paragraph{Closed system.}
  Figure \ref{subfig:LamKus02-0_CM} provides another validity check,
  showing that the conserved moieties are constant; because the reduced 
  order equations (\S~\ref{app:reduc-order-equat}) were used, this constraint is
  automatically enforced and thus numerical drift is avoided.
Figure \ref{subfig:LamKus02-0_FoutATPase} shows mass flows within the system.
\paragraph{Open system.}
  Figure \ref{subfig:LamKus02-4_X} shows the evolution of simulated
  concentrations for $ATP$, $ADP$, $P$, $GLY$ and $LAC$ which reach
  new steady state values due to flows induced by the $ATPase$ reaction.
  As with Figure \ref{subfig:LamKus02-0_CM}, Figure
  \ref{subfig:LamKus02-4_CM} shows the conserved moieties are
  constant for the open system.
  In contrast to Figure \ref{subfig:LamKus02-0_FoutATPase}, Figure
  \ref{subfig:LamKus02-4_FoutATPase} shows mass flow which, in this
  open-system context settle to non-zero values.




\subsection{Virtual Reference Environment}
\label{sec:virt-refer-envir}
The software required to generate all of the simulation figures shown
in this paper from the bond graph representation is packaged in the
form of a \emph{Virtual Reference Environment} \citep{HurBudCra14}. It
is available at \url{https://sourceforge.net/projects/hbgm/}.


\end{document}